\documentclass[aps,prl,reprint,superscriptaddress,nofootinbib,floatfix,longbibliography]{revtex4-2}

\usepackage{amsmath,amssymb,bm}
\usepackage{graphicx}
\usepackage{booktabs}
\usepackage{siunitx}
\usepackage{hyperref}
\hypersetup{hidelinks}
\usepackage{orcidlink}
\newcommand{\rev}[1]{\textcolor{black}{#1}}

\newcommand{\bk}{\bm{k}}

\newcommand{\sgn}{\operatorname{sgn}}
\newcommand{\Tt}{\mathcal{A}}
\newcommand{\ii}{\mathrm{i}}

\begin{document}

\title{Winding-Controlled Topological Descendants of Antialtermagnets}

\author{Sahar Izadi Vishkayi\orcidlink{0000-0002-0314-1007}}
\email{izadi.123@gmail.com}
\affiliation{International Research Centre MagTop, Institute of Physics, Polish Academy of Sciences, Aleja Lotnik\'ow 32/46, PL-02668 Warsaw, Poland}

\author{Carmine Autieri\orcidlink{0000-0002-5008-8165}}
\email{autieri@magtop.ifpan.edu.pl}
\affiliation{International Research Centre MagTop, Institute of Physics, Polish Academy of Sciences, Aleja Lotnik\'ow 32/46, PL-02668 Warsaw, Poland}

\author{Meysam Bagheri Tagani\orcidlink{0000-0001-7065-3195}}
\email{mtagani@magtop.ifpan.edu.pl}
\affiliation{International Research Centre MagTop, Institute of Physics, Polish Academy of Sciences, Aleja Lotnik\'ow 32/46, PL-02668 Warsaw, Poland}
\affiliation{Department of Physics, University of Guilan, P. O. Box 41335-1914, Rasht, Iran}
\date{\today}

\begin{abstract}
Odd-parity spin splitting constrains magnetic band crossings but
does not determine their topological descendants. We show that the
Chern contribution of an isolated inverted sector is controlled by
the phase winding of its transverse mass pair: when the masses and
tangential velocities reverse together, contributions from
successive Dirac points add rather than cancel. This distinguishes
topological winding from the polynomial degree labeling the magnetic
parent. For specified complementary masses, periodic tetragonal
$h$- and hexagonal $j$-wave models host eight and twelve
same-chirality Weyl points on a momentum plane, with gapped-slice
Chern numbers $|C|=4$ and $6$, respectively. A microscopic
four-band $p$-wave model identifies an antiunitary obstruction to
first-order transverse masses and demonstrates a symmetry-broken
$C=1$ phase at quarter filling. First-principles calculations for
the proposed coplanar spin-vortex texture of LaMnAu$_5$ reveal
the $h$-wave nodal pattern, its anisotropic spin--orbit reconstruction,
and a mirror-channel spectral response that reverses under
opposite spin rotations. Together, these results distinguish
the symmetry of an odd-parity magnetic parent from the mass
structure controlling the topology of its descendants.
\end{abstract}

\maketitle

\paragraph*{\rev{Introduction.---}}
\rev{Altermagnets exhibit even-in-momentum
splitting~\cite{Hayami2020BottomUp,Smejkal2022PRX,Smejkal2022Landscape}, whereas noncollinear
antiferromagnets can support non-relativistic odd-parity spin
momentum locking in both preserved~\cite{Hayami2020Antisymmetric,Hellenes2023Pwave,Yu2025OddParity}
and broken time-reversal symmetry\cite{shen2026roomtemperatureferroelectricpwavemagnet}.
Other effective p-wave magnets have been proposed as a combination of multiple spin-momentum lockings\cite{Fakhredine25b,pvpj-dbnq}.
Such states are already accessible experimentally. In spiral NiI$_2$,
the $p$-wave spin polarization is coupled to electrically switchable
magnetic chirality~\cite{Song2025Switching}; a commensurate helical
$p$-wave metal shows a large anomalous Hall response in the presence of a weak net magnetization~\cite{Yamada2025Nature}. Recent studies have explored the topological properties of p-wave magnets\cite{reja2026bicircularlighttunabletopological}; however, several open questions remain. 
These findings raise a question beyond identifying odd-parity splitting:
what determines the topology of the bands when their nodes are reconstructed?}

\rev{Anti-spin-Laue and spin-point-group classifications now identify
symmetry-allowed spin-splitting forms, including high-order $h$ and $j$
waves~\cite{Lange2026ASLG,Elcoro2026HighOrder}. Such classifications
locate the parent nodal structure, but do not specify the masses that
can gap it. We use the term topological descendant for a Chern or Weyl phase that emerges when additional symmetry-breaking terms reconstruct the nodal bands of an odd-parity magnetic parent. Floquet $f$-wave Chern phases and odd-parity magnetic Chern
insulators demonstrate possible topological
outcomes~\cite{Zhu2026Floquet,LinVila2026OddParity}.
The unresolved issue for a multiply nodal parent is when the contributions
from its crossings reinforce rather than cancel. The relevant quantity
need not be the polynomial degree used to label the spin splitting.}

\rev{Here we formulate a criterion in terms of the winding of the
\emph{projected transverse mass pair} around a band-inversion contour.
At successive Dirac points, the mass and tangential velocity can change
sign together, making their Berry-curvature contributions add coherently.
For an isolated, properly regularized inverted sector, the resulting
Chern number is fixed by that mass winding. The distinction is pronounced
for tetragonal $h$ and hexagonal $j$ forms: their minimum polynomial
degrees are five and seven, while the corresponding planar angular orders
are four and six. With specified complementary masses, periodic models
then realize eight and twelve same-chirality Weyl points, respectively,
and gapped momentum slices with $|C|=4$ and $6$.
These charges belong to the constructed descendants, not to every magnet
bearing the same wave label~\cite{SupplementalMaterial}.}

\rev{To establish what symmetry breaking is needed, we project the complete
internal matrix basis of a microscopic four-band $p$-wave model.
The calculation identifies six first-order transverse mass channels, none of which
can preserve both protecting antiunitary symmetries; complementary
perturbations yield a lowest-band $C=1$ phase.
For a material test, we consider the coplanar spin-vortex structure
proposed for LaMnAu$_5$~\cite{Siebeneichler2024LaMnAu5}, already classified
as an $h$-wave candidate~\cite{Elcoro2026HighOrder,Lange2026ASLG}.
First-principles calculations resolve its nodal spin splitting and show
that spin-orbit coupling (SOC) produces an anisotropic spectral reconstruction that reverses
under opposite in-plane spin rotations. Thus the material calculation
establishes control of the magnetic parent, while the model analysis
identifies the additional mass winding required for a topological descendant.}

\begin{figure*}[t]
\centering
\includegraphics[width=0.98\textwidth]{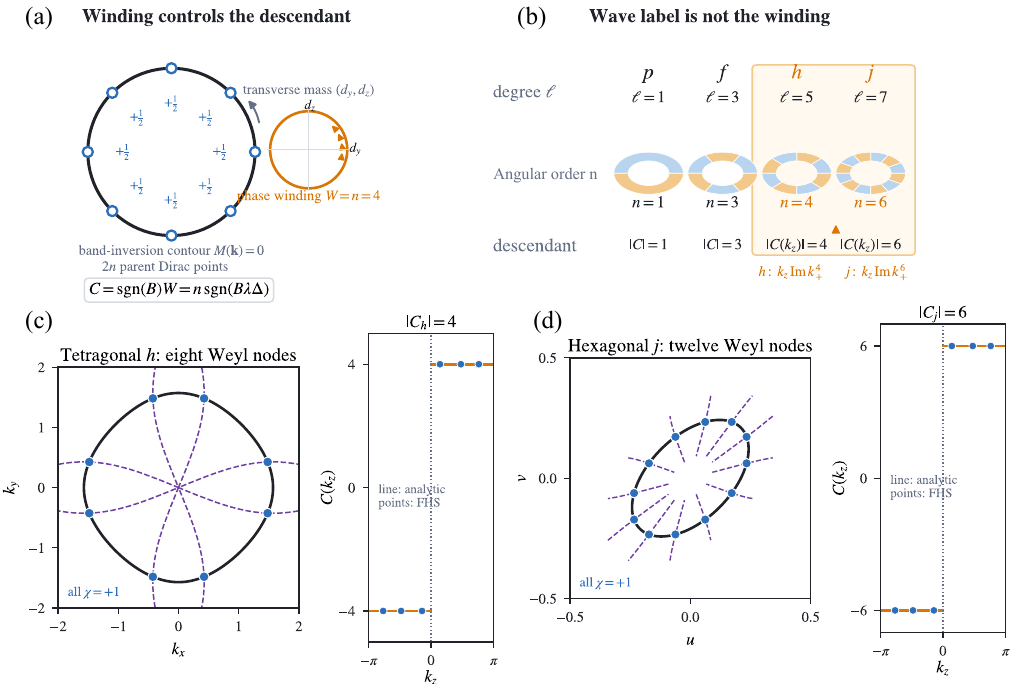}
\caption{\rev{\textbf{Mass winding and topological descendants.}
(a) A representative $n=4$ inversion contour illustrates the general
$2n$ Dirac crossings of a sine harmonic. For the specified cosine
partner, alternating Dirac masses and tangential velocities give
coherent half-Chern contributions; the transverse pair winds around
the origin.
(b) The minimum polynomial degree $\ell$ labels the parent splitting,
whereas the planar angular order $n$ characterizes the harmonic.
The indicated Chern numbers refer to specified complementary mass
pairs and are not fixed by $\ell$ alone.
(c,d) Exactly periodic tetragonal $h$- and hexagonal $j$-wave models
have eight and twelve same-chirality Weyl points on $k_z=0$ and
$|C(k_z)|=4$ and $6$ on gapped slices. Opposite monopole charges
occur at $k_z=\pi$. Curves and symbols denote analytical and
Fukui--Hatsugai--Suzuki (FHS) results; further checks are in Figs.~S1--S3.}}
\label{fig:winding}
\end{figure*}

\paragraph*{\rev{Chern topology from mass winding.---}}
\rev{We first examine how a gapped nodal structure contributes to Chern
topology in an inverted two-band sector. Let $\bk=(k_x,k_y)$, $k=|\bk|$, and
$\theta$ be the polar angle. With Pauli matrices $\mu_i$ acting in the
effective subspace, consider}
\begin{equation}
\begin{aligned}
 h_n(\bk)={}&M(k)\mu_x+\lambda k^n\cos(n\theta)\mu_y\\
 &+\Delta k^n\sin(n\theta)\mu_z,\qquad M(k)=m-Bk^2 .
\end{aligned}
\label{eq:planar}
\end{equation}
\rev{The sine harmonic sets the parent nodal directions, while the cosine
component supplies a complementary transverse term. Both the inversion
term and this second component are specified ingredients of the model;
they do not follow from the magnetic wave label alone. The integer $n$
is the in-plane angular order: odd $n$ describes a planar odd-parity
harmonic, while even $n$ will appear on fixed-$k_z$ slices of
three-dimensional odd-parity forms.}

\rev{For $mB>0$ and $\lambda=0$, the inversion contour $M(k)=0$ has radius
$k_0=\sqrt{m/B}$. Its intersections with $\sin(n\theta)=0$ give $2n$
Dirac points at $\theta_a=a\pi/n$. Once $\lambda\ne0$, the mass at
successive points alternates as $(-1)^a$, but so does the tangential
Dirac velocity. Their signs therefore compensate, and each gapped
crossing contributes $\tfrac12\sgn(B\lambda\Delta)$ to the lower-band
Chern number [Fig.~\ref{fig:winding}(a)]~\cite{Sticlet2012Chern}.}

\rev{The same result has a simple geometric interpretation. Along the
inversion contour, the transverse components
$d_y=\lambda k^n\cos(n\theta)$ and $d_z=\Delta k^n\sin(n\theta)$ trace
a closed curve in the $(d_y,d_z)$ plane. When $\lambda\Delta\ne0$,
its phase winds $n$ times, with orientation fixed by the signs of the
two coefficients. Thus, for a gapped periodic completion in which no
other band inversion contributes,}
\begin{equation}
\begin{aligned}
 W&=\frac{1}{2\pi}\oint_{M=0}d\arg[d_y(\bk)+\ii d_z(\bk)],\\
 C&=\sgn(B)W=n\,\sgn(B\lambda\Delta).
\end{aligned}
\label{eq:winding}
\end{equation}
\rev{The parent nodal pattern alone does not determine $W$: a different
transverse partner can give a different winding or zero winding. Nor
does the local Dirac-point count automatically determine a global
Chern number. In the exact triangular-lattice $f$-wave example, the
first band inversion gives $C=3$, whereas a second inversion changes
it to $C=-6$. Periodic lattice calculations, Berry-curvature
integration, and Wilson-loop checks are given in the SM
(Figs.~S1 and S2)~\cite{Fukui2005,SupplementalMaterial}.}

\paragraph*{\rev{High-wave descendants.---}}
\rev{The spin-point-group classification identifies coplanar tetragonal
and hexagonal odd-parity splittings whose leading momentum dependences are~\cite{Elcoro2026HighOrder}}
\begin{align}
 \Phi_h&\propto k_xk_yk_z(k_x^2-k_y^2)
 =\frac{k_z}{4}\operatorname{Im}(k_x+\ii k_y)^4,
 \label{eq:hform}\\
 \Phi_j&\propto k_xk_yk_z(3k_x^2-k_y^2)(k_x^2-3k_y^2)
 =\frac{k_z}{2}\operatorname{Im}(k_x+\ii k_y)^6.
 \label{eq:jform}
\end{align}
\rev{Their minimum polynomial degrees are five and seven, whereas their
in-plane angular orders at fixed nonzero $k_z$ are four and six. The
factor $k_z$ makes the full splitting odd in momentum. These real
harmonics determine nodal directions, but not a Chern number: a second,
transverse component is still needed. We construct periodic descendants
by specifying the corresponding cosine quadrature, without assuming
that it is required by the parent magnetic symmetry.}

\rev{For the tetragonal case, let $z=\sin k_x+\ii\sin k_y$ and take}
\begin{equation}
\begin{aligned}
 H_h={}&[m+\cos k_x+\cos k_y]\mu_x
 +\lambda\operatorname{Re}(z^4)\mu_y\\
 &+\Delta\sin k_z\operatorname{Im}(z^4)\mu_z .
\end{aligned}
\label{eq:hreg}
\end{equation}
\rev{Since $z(C_4\bk_\parallel)=\ii z(\bk_\parallel)$, both transverse
harmonics respect the fourfold rotation. For $-2<m<0$ and
$\lambda\Delta\sin k_z\ne0$, each two-dimensional slice is gapped,
with $C_h(k_z)=4\,\sgn(\lambda\Delta\sin k_z)$.
At $k_z=0$, the odd transverse component vanishes and eight Weyl points
remain where the inversion contour meets $\operatorname{Re}(z^4)=0$.
For the parameters shown in Fig.~\ref{fig:winding}(c), each carries
charge $+1$; the eight nodes at $k_z=\pi$ have the opposite charge.}

\rev{For the hexagonal case, we use three $120^\circ$-separated triangular
bonds $\bm\delta_a$ to define
$g=\sum_{a=1}^3\omega^{a-1}\sin(\bk_\parallel\cdot\bm\delta_a)$ and
$\Phi=\sum_a[1-\cos(\bk_\parallel\cdot\bm\delta_a)]$, where
$\omega=e^{2\pi\ii/3}$. The exact relation
$g(C_6\bk_\parallel)=e^{\ii\pi/3}g(\bk_\parallel)$ makes $g^6$
sixfold invariant. The corresponding model is}
\begin{equation}
 H_j=[m-B\Phi]\mu_x+\lambda\operatorname{Re}(g^6)\mu_y
 +\Delta\sin k_z\operatorname{Im}(g^6)\mu_z .
\label{eq:jreg}
\end{equation}
\rev{In its first inverted regime, $0<m/B<4$, the gapped slices have
$C_j(k_z)=6\,\sgn(B\lambda\Delta\sin k_z)$, with twelve same-chirality
Weyl points at $k_z=0$ and twelve opposite-charge partners at $k_z=\pi$
[Fig.~\ref{fig:winding}(d)]. These compensating charges satisfy lattice
Weyl-node neutrality~\cite{Wan2011Weyl,Armitage2018Weyl}.
The results establish topological phases of the \emph{specified}
rotationally compatible Hamiltonians; identifying the corresponding
mass operators in a material requires its relativistic magnetic
symmetries and an isolated electronic subspace~\cite{SupplementalMaterial}.}

\begin{figure*}[t]
\centering
\includegraphics[width=0.98\textwidth]{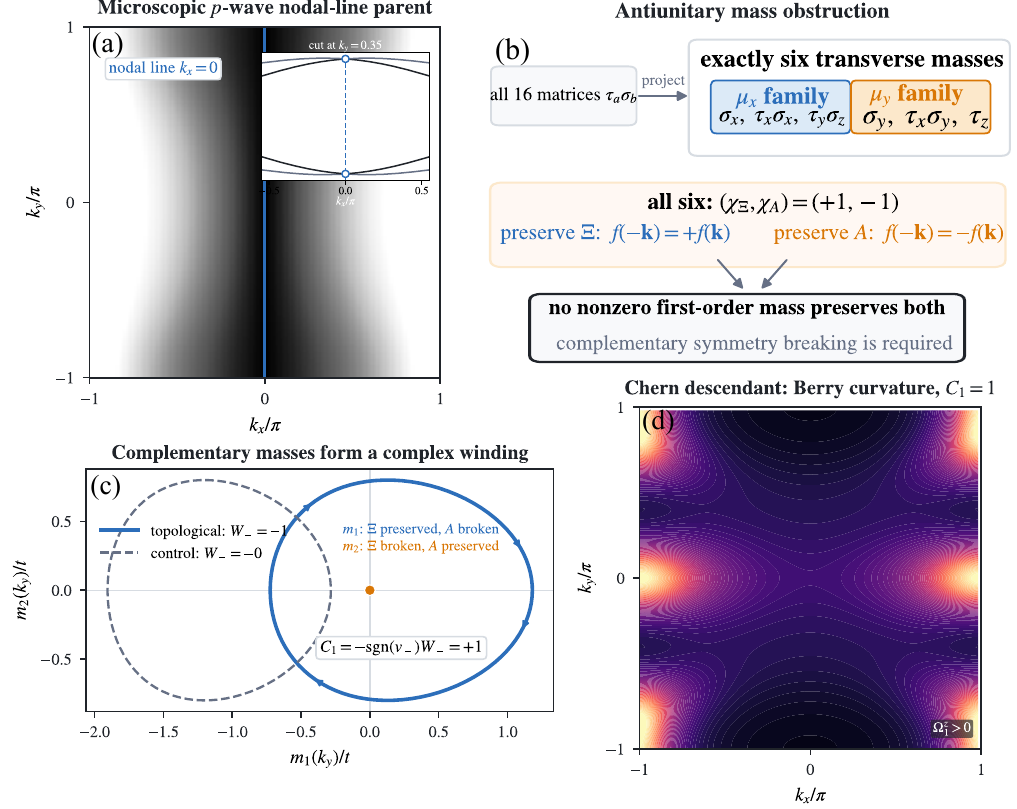}
\caption{\rev{\textbf{Microscopic constraints on $p$-wave topology.}
(a) Two nodal doublets of the four-band parent along $k_x=0$.
(b) Six of the 16 internal Hermitian matrices generate first-order
transverse masses; every one breaks at least one of the two protecting
antiunitary symmetries.
(c) Two symmetry-distinct perturbations produce either a winding
($W_-=-1$) or a nonwinding ($W_-=0$) projected mass loop.
(d) Berry curvature of the isolated lowest band in the topological
reference model, with $C_1=1$ at quarter filling. Complete projections,
gap conditions, and independent Chern checks appear in Fig.~S4.}}
\label{fig:micro}
\end{figure*}

\paragraph*{\rev{Microscopic symmetry constraints.---}}
\rev{The preceding lattice models specify transverse masses. To determine
how a microscopic magnetic Hamiltonian restricts such terms, we consider
a four-band $p$-wave parent based on established tight-binding
models~\cite{Hellenes2023Pwave,Chakraborty2025Edelstein},}
\begin{align}
H_p(\bk)=&\;2t\cos\frac{k_x}{2}\tau_x+2t_y\cos k_y\,\bf{1}
+2J\sin\frac{k_x}{2}\tau_y\sigma_x\notag\\
&+[\delta_0+2J\cos k_y]\tau_z\sigma_y .
\label{eq:pmodel}
\end{align}
\rev{The Pauli matrices $\tau_i$ and $\sigma_i$ act on the two-site and
spin spaces. The parameters $t_y$ and $\delta_0$ retain the relevant
symmetries while allowing us to isolate the nodal line. The parent has
two effective antiunitary symmetries,
$\Xi=\ii\sigma_xK$ and $\Tt=\tau_x\ii\sigma_yK$, which reverse $\bk$,
and a unitary symmetry $U_y=\ii\sigma_y$ taking
$(k_x,k_y)\to(-k_x,k_y)$. Here $K$ denotes complex conjugation.
These operations are symmetries of the Bloch model, not a claim that
the ordered moments preserve bare time reversal~\cite{Luo2026Hidden}.
Their product gives a fixed-momentum symmetry
$\mathcal S=\tau_x\sigma_z$, up to a phase, with
$\{\mathcal S,U_y\}=0$. On $k_x=0$, this enforces two nodal
doublets [Fig.~\ref{fig:micro}(a)].}

\rev{Projecting the full set of 16 Hermitian $\tau_a\sigma_b$
matrices onto either doublet yields six nonzero transverse masses at
first order in a perturbation:}
\begin{equation}
\begin{split}
\mu_x:&\quad \sigma_x,\;\tau_x\sigma_x,\;\tau_y\sigma_z,\\
\mu_y:&\quad \sigma_y,\;\tau_x\sigma_y,\;\tau_z.
\end{split}
\label{eq:sixmass}
\end{equation}
\rev{Every matrix in this list is even under $\Xi$ and odd under $\Tt$.
Consequently, a real coefficient $f(\bk)$ could preserve both
antiunitaries only if it were simultaneously even and odd under
$\bk\to-\bk$. No nonzero first-order transverse mass can meet both
conditions [Fig.~\ref{fig:micro}(b)]. This obstruction is exact within
the specified four-band Hilbert space at the parent nodal line.}

\rev{A concrete pair of perturbations is
$H_{\rm r}=\lambda\sin k_y\,\sigma_y$ and
$H_{\rm b}=(M_0+M_y\cos k_y)\sigma_x$.
The former breaks $\Xi$ but preserves $\Tt$ and $U_y$, whereas
the latter preserves $\Xi$ but breaks $\Tt$ and $U_y$; together
they remove the antiunitary and reflection constraints that would
forbid a nonzero Chern number. Projected onto the lower doublet, they produce
a complex mass $m_1(k_y)+\ii m_2(k_y)$ that winds once when
$|M_0|<|M_y|$, with $W_-=-1$ for our sign convention, and does not
wind when $|M_0|>|M_y|$ [Fig.~\ref{fig:micro}(c)]. The line contribution
is $C_{\rm line}=-\sgn(v_-)W_-$, where $v_-$ is the normal band velocity.
For the reference parameters, a sewing-aware calculation of the full
four-band Hamiltonian gives $(C_1,C_2,C_3,C_4)=(+1,-1,+1,-1)$.
The isolated lowest band therefore has $C_1=1$ at quarter filling,
confirmed by Kubo, FHS, and Wilson-loop results [Fig.~\ref{fig:micro}(d)
and Fig.~S4]. The first two bands together have zero Chern number.
The perturbations are specified symmetry channels, not a fitted
SOC or exchange field for a particular material. Experimentally, SOC and weak magnetization have been linked to a
large Hall response in a metallic $p$-wave magnet~\cite{Yamada2025Nature},
although that observation does not identify our specific mass terms.}

\begin{figure*}[t]
\centering
\includegraphics[width=0.99\textwidth]{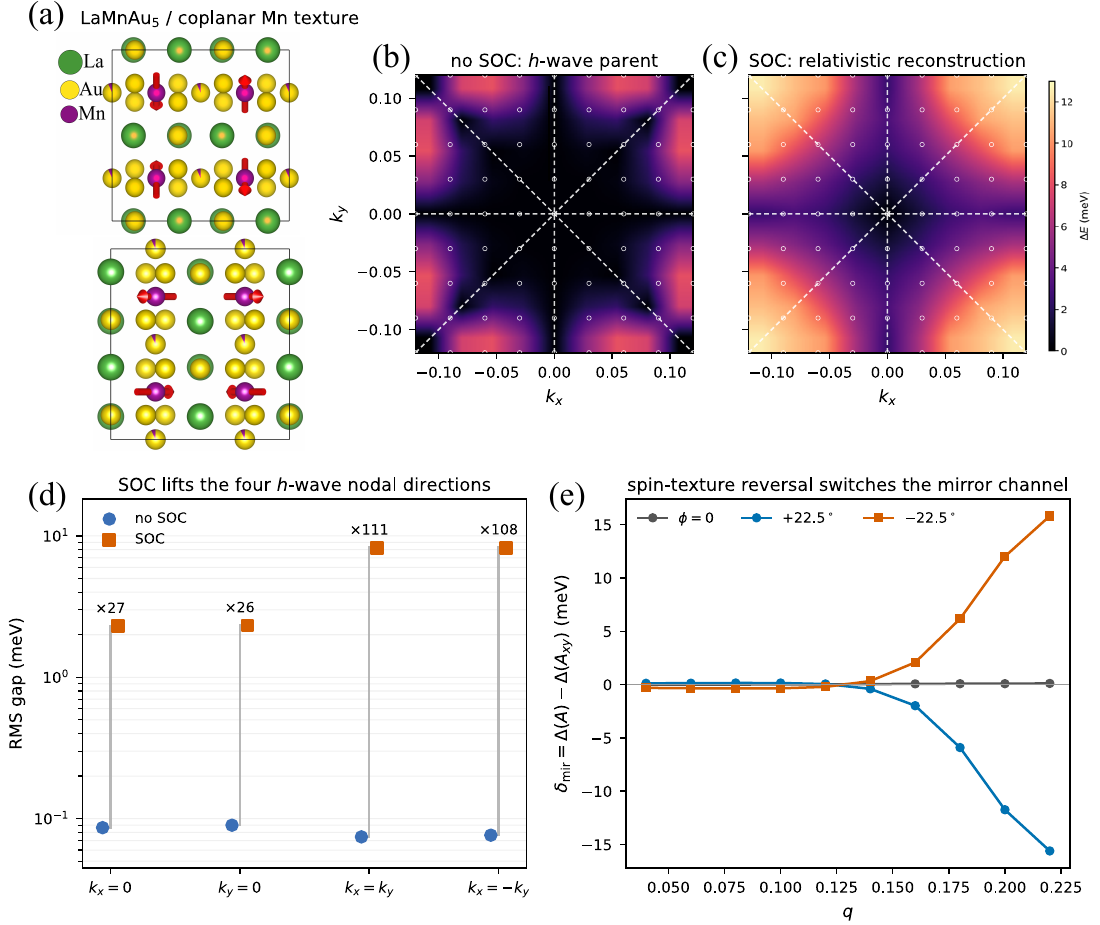}
\caption{\rev{\textbf{Nonrelativistic $h$-wave parent and its relativistic
reconstruction in LaMnAu$_5$.}
(a) Crystal structure and the coplanar Mn SVC texture, viewed in the
$x$--$z$ and $y$--$z$ planes. (b,c) Low-energy doublet splitting without
and with SOC at fixed $k_z=0.10$, displayed on the same momentum grid
and energy scale. Dashed lines indicate the four parent nodal directions;
open circles identify sampled points. 
(d) RMS splitting along the four parent nodal directions, showing the
strong, direction-dependent relativistic reconstruction.
(e) Splitting difference between mirror-related momentum paths for the
reference SVC and for coherent in-plane rotations by $\pm22.5^\circ$.
Opposite rotations reverse this spectral anisotropy. Band tracking,
symmetry checks, and independent spectral controls are documented in
Figs.~S5--S7.}}
\label{fig:material}
\end{figure*}

\paragraph*{\rev{Material test: LaMnAu$_5$.---}}
\rev{We examine the relativistic reconstruction of an $h$-wave parent in
metallic LaMnAu$_5$, which contains frustrated Mn square nets. Neutron
diffraction permits a coplanar hedgehog spin-vortex-crystal (SVC)
structure and a competing collinear refinement, while relativistic
calculations favor the SVC~\cite{Siebeneichler2024LaMnAu5}. We adopt the
reported SVC [Fig.~\ref{fig:material}(a)], whose $h$-wave character
has been identified by spin-symmetry
classification~\cite{Elcoro2026HighOrder,Lange2026ASLG}, and investigate
its electronic response to SOC and magnetic orientation.}

\rev{Without SOC, the splitting of a low-energy doublet is strongly
suppressed along $k_x=0$, $k_y=0$, and $k_x=\pm k_y$ at fixed nonzero $k_z$
[Fig.~\ref{fig:material}(b)]. These are the four nodal directions of
the $h$-wave form $k_xk_yk_z(k_x^2-k_y^2)$, and the calculated angular
dependence follows its leading behavior. Independent orbital- and
spin-character tracking checks the doublet identification through
neighboring-band reorderings (Fig.~S5).}

\rev{SOC reconstructs the same low-energy sector anisotropically
[Fig.~\ref{fig:material}(c,d)]. The splitting along the parent nodal
directions rises to a few meV on the axes and is substantially larger
on the diagonals. The result is not a uniform additional separation:
relativistic coupling changes the angular pattern of the electronic
bands. Matched momentum paths and band-alignment checks are given in
Fig.~S6.}

\rev{The effect also depends on the orientation of the SVC. We compare
the reference texture with coherent in-plane rotations of all Mn moments
by $\phi=\pm22.5^\circ$. For paths $A$ and $A_{xy}$ related by
$k_x\leftrightarrow k_y$, define
$\delta_\Delta(\phi,q)=\Delta_A(\phi,q)-\Delta_{A_{xy}}(\phi,q)$.
This difference is small for the reference state, but opposite rotations
produce nearly equal and opposite mirror-channel responses
[Fig.~\ref{fig:material}(e)]. The band-pair midpoint shows the same
reversal independently (Fig.~S7). LaMnAu$_5$ thus demonstrates a
magnetic-orientation-dependent relativistic reconstruction of a
symmetry-classified odd-parity parent. The sensitivity to magnetic
orientation also recalls the role of SOC and additional magnetic
symmetry breaking in the anomalous Hall response of helical $p$-wave
magnets~\cite{Yamada2025Nature,Okumura2025Spiral}.}

\paragraph*{\rev{Discussion.---}}
\rev{The central distinction is between the symmetry label of an
odd-parity magnetic parent and the winding of the transverse mass pair
that reconstructs its nodal bands. The latter fixes the topological
contribution of an isolated inverted sector. Our periodic $h$- and
$j$-wave constructions consequently realize Weyl descendants with
slice Chern numbers $|C|=4$ and $6$ for specified mass partners, rather
than identifying polynomial degrees five and seven with topological
invariants. The microscopic $p$-wave model further shows how
complementary symmetry-breaking channels remove the antiunitary
obstruction and generate a lowest-band $C=1$ phase.}

\rev{LaMnAu$_5$ provides a material counterpart at the level of the
magnetic parent: its $h$-wave nodal fingerprint, anisotropic SOC
reconstruction, and opposite-rotation spectral response are resolved
by first-principles calculations. The model Chern numbers belong to
the explicitly constructed mass pairs, while the material results
establish how a real odd-parity electronic structure responds to
relativistic symmetry breaking. The resulting materials strategy is
clear: classify the parent, identify allowed transverse terms, and
evaluate their winding in an isolated band subspace.}

\begin{acknowledgments}
M.B.T. acknowledges support from the Polish National Agency for Academic Exchange (NAWA) under the ULAM program, project BPN/ULM/2025/1/00156/U/00001, and from the Iran National Science Foundation under project 40506767. This work was supported by the Foundation for Polish Science Project ``MagTop,'' FENG.02.01-IP.05-0028/23, co-financed by the European Union under the European Funds for a Smart Economy Program 2021--2027. We acknowledge computing resources of the Interdisciplinary Centre for Mathematical and Computational Modelling, University of Warsaw (grants g91-1418, g91-1419, g96-1808, g96-1809, and g103-2540), and the Poznan Supercomputing and Networking Center (grants pl0267-01, pl0365-01, pl0807, and pl0471-01).
\end{acknowledgments}

\paragraph*{Data availability.---}
The numerical data and custom analysis scripts supporting the figures are
available from the author upon reasonable request.  A persistent public
archive containing the figure-generation scripts with both the raw and processed
first-principles audit data will be supplied at publication.

\paragraph*{AI-use disclosure.---}
AI tools were used to assist with language editing,  and preparation of  plotting scripts.  The authors
reviewed the code and numerical outputs and are responsible for the scientific
content, interpretation, and conclusions.

\nocite{Yu2011Wilson,Kresse1996VASP,Kresse1999PAW,Perdew1996PBE,Roig2024PRB,Sim2026Quantum,Leeb2026Collinear,Kravchuk2026Magnon,Li2026Orbital}
\bibliography{references}

\section*{End Matter: Numerical and first-principles protocol}
\label{sec:endmatter}
The topological calculations use the Berry-curvature convention
$\Omega_n^z=-2\operatorname{Im}\sum_{m\neq n}
\langle n|v_x|m\rangle\langle m|v_y|n\rangle/(E_m-E_n)^2$.
Chern numbers are evaluated independently from direct interband curvature and
the lattice-gauge FHS construction~\cite{Fukui2005}; Wilson phases are closed
with the appropriate orbital sewing matrix in the four-band model.  Weyl
chiralities are obtained from the determinant of the linearized three-
component $\bm d$-map.  Figs.~S1--S4 document sign conventions, mesh checks,
phase boundaries, and finite-parameter tests.

First-principles calculations were performed with the projector-augmented-wave
implementation of density-functional theory in VASP
\cite{Kresse1996VASP,Kresse1999PAW}, using the PBE generalized-gradient
approximation~\cite{Perdew1996PBE}.  The production workflow uses a
$550$-eV plane-wave cutoff, noncollinear magnetism with spatial symmetry
disabled, and the coplanar SVC directions derived from the reported
LaMnAu$_5$ magnetic structure.    The SOC-rotation controls are obtained by rotating
all Mn moments coherently by $\phi=\pm22.5^\circ$ in the coplanar plane. The Brillouin zone was sampled with a $\mathbf{k}$-point mesh density of $2\pi \times 0.04\text{ \AA}^{-1}$, with an energy convergence threshold of $10^{-6}\text{ eV}$.
The complete first-principles protocol, raw-grid definitions, band-tracking
criteria, and numerical symmetry residuals are given in the SM.

\end{document}


\title{Supplemental Material for ``Winding-Controlled Topological Descendants of Anti-Altermagnets''}

\author{Sahar Izadi Vishkayi\orcidlink{0000-0002-0314-1007}}
\email{izadi.123@gmail.com}
\affiliation{International Research Centre MagTop, Institute of Physics, Polish Academy of Sciences, Aleja Lotnik\'ow 32/46, PL-02668 Warsaw, Poland}

\author{Carmine Autieri\orcidlink{0000-0002-5008-8165}}
\email{autieri@magtop.ifpan.edu.pl}
\affiliation{International Research Centre MagTop, Institute of Physics, Polish Academy of Sciences, Aleja Lotnik\'ow 32/46, PL-02668 Warsaw, Poland}

\author{Meysam Bagheri Tagani\orcidlink{0000-0001-7065-3195}}
\email{mtagani@magtop.ifpan.edu.pl}
\affiliation{International Research Centre MagTop, Institute of Physics, Polish Academy of Sciences, Aleja Lotnik\'ow 32/46, PL-02668 Warsaw, Poland}
\affiliation{Department of Physics, University of Guilan, P. O. Box 41335-1914, Rasht, Iran}

\date{\today}
\maketitle

\section{Scope, terminology, and conventions}
\label{sec:scope}

\rev{This Supplemental Material gives the model definitions, derivations,
numerical checks, and first-principles analysis used in the Letter.
The model calculations establish topological invariants for specified
Hamiltonians. The LaMnAu$_5$ calculations establish the nonrelativistic
parent and its SOC-dependent spectrum; they do not assign a material Chern
number or Weyl charge.}

\rev{We distinguish three quantities. The wave label $\ell$ is the minimum
polynomial degree of the symmetry-allowed spin splitting. The integer $n$
is the angular order of a planar harmonic. The signed integer $W$ is the
phase winding of a nonvanishing complex mass pair. A real spin-splitting
polynomial determines neither a second mass component nor its winding.
The parent classification is taken from
Refs.~\cite{Lange2026ASLG,Elcoro2026HighOrder}; the matching mass partners
in our $h$- and $j$-wave models are specified explicitly.}

\rev{The microscopic $p$-wave analysis uses the effective antiunitary symmetries
of the Bloch Hamiltonian. These are not identified with microscopic time
reversal of the magnetic order~\cite{Luo2026Hidden}.
Relevant microscopic and symmetry studies are
Refs.~\cite{Hellenes2023Pwave,Brekke2024PRL,Chakraborty2025Edelstein,Yu2025OddParity,Mitscherling2026Microscopic,Hirschmann2026Fwave}.
Odd-parity structure also occurs in magnon spectra, as discussed separately
in Ref.~\cite{Kravchuk2026Magnon}.}

\rev{The Berry curvature convention is}
\begin{equation}
\Omega_n^z(\bk)=-2\,\mathrm{Im}\sum_{m\neq n}
\frac{\langle n|v_x|m\rangle\langle m|v_y|n\rangle}
{(E_m-E_n)^2},
\label{eq:Sberry}
\end{equation}
\rev{Here $v_i=\partial H/\partial k_i$ in model units with $\hbar=1$.
This is the standard interband convention~\cite{Xiao2010Berry}.
For a two-level Hamiltonian $h=\bm d\cdot\bm\mu$, it gives}
\begin{equation}
\Omega_-^z=\frac12\hat{\bm d}\cdot
(\partial_{k_x}\hat{\bm d}\times\partial_{k_y}\hat{\bm d}).
\label{eq:Stwolevel}
\end{equation}
\rev{Chern signs refer to the lower band and the orientation $(k_x,k_y)$.
Weyl charge uses the same lower-band convention with an outward-oriented
surface in $(k_x,k_y,k_z)$. Model momenta are dimensionless in the
corresponding lattice units. FHS denotes the Fukui--Hatsugai--Suzuki
lattice-gauge method~\cite{Fukui2005}; Wilson loops provide an independent
check~\cite{Yu2011Wilson}. The mass winding is evaluated in the specified
effective basis, with the boundary sewing retained when required.}

\rev{Our lattice-gauge orientation is matched to the Berry-curvature
convention of Eq.~\eqref{eq:Sberry}. For forward overlap links
$U_a(\bk)=\langle u(\bk)|u(\bk+\delta_a)\rangle/
|\langle u(\bk)|u(\bk+\delta_a)\rangle|$,
the plaquette argument enters the Chern number with a minus sign.
Equivalently, backward overlap links give the positive orientation.
This convention fixes the signs quoted for the Weyl charges and slice
Chern numbers.}

\section{\rev{Planar winding criterion and coherent Dirac contributions}}
\label{sec:planar}

\rev{Let $\bk=(k_x,k_y)$, $k=|\bk|$, and $\theta$ be its polar angle.
Consider the real harmonic}
\begin{equation}
 h_n^{(0)}(\bk)=\Delta k^n\sin(n\theta)\mu_z,
 \qquad n=1,2,\ldots,
\label{eq:Snparent}
\end{equation}
\rev{where $n$ denotes angular order, not a topological invariant.
Odd $n$ gives a planar odd-parity harmonic. Even $n$ is used below on
fixed-$k_z$ slices of three-dimensional forms. We embed this component in
an inverted two-band model and add the matching cosine component:}
\begin{equation}
 h_n=(m-Bk^2)\mu_x+\lambda k^n\cos(n\theta)\mu_y
 +\Delta k^n\sin(n\theta)\mu_z .
\label{eq:Sndesc}
\end{equation}
\rev{The $\mu_i$ are Pauli matrices in the effective subspace. The inversion
term and the quadrature are assumptions of this construction.
For $mB>0$ and $\lambda=0$, the bands meet where $d_x=d_z=0$:}
\begin{equation}
 k_0=\sqrt{m/B},\qquad \theta_a=\frac{a\pi}{n},
 \quad a=0,\ldots,2n-1.
\label{eq:Sroots}
\end{equation}
\rev{Introduce radial momentum $q_r$ and tangential momentum
$q_t=k_0\delta\theta$. The frame $(\hat{\bm e}_r,\hat{\bm e}_t)$ has the
same orientation as $(k_x,k_y)$. Near crossing $a$,}
\begin{equation}
 h_a=-2Bk_0q_r\mu_x+\lambda k_0^n(-1)^a\mu_y
 +n\Delta k_0^{n-1}(-1)^a q_t\mu_z .
\label{eq:SlocalDirac}
\end{equation}
\rev{The mass and tangential velocity both reverse sign between neighboring
crossings. Their product therefore keeps the same sign. Using the
standard massive-Dirac contribution~\cite{Sticlet2012Chern} with
Eq.~\eqref{eq:Stwolevel}, each cone contributes}
\begin{equation}
 c_a=\frac12\sgn(B\lambda\Delta),
\label{eq:Shalf}
\end{equation}
\rev{independent of $a$. In a periodic completion with no additional topological
sector, the total is}
\begin{equation}
 C=\sum_{a=0}^{2n-1}c_a
 =n\,\sgn(B\lambda\Delta).
\label{eq:SCn}
\end{equation}
\rev{Equivalently, define the winding on the counterclockwise inversion contour,
where the transverse pair must not vanish:}
\begin{equation}
 W=\frac{1}{2\pi}\oint_{M=0}d\arg[d_y(\bk)+\ii d_z(\bk)] .
\label{eq:SW}
\end{equation}
\rev{For the radial mass orientation in Eq.~\eqref{eq:Sndesc},
$C=\sgn(B)W$. This relation uses the \emph{actual} pair $d_y+\ii d_z$.
A different transverse component can give a different winding while
leaving the parent sine harmonic unchanged. The four-band model in
Sec.~\ref{sec:pwave} provides an explicit winding/nonwinding comparison
within one family of symmetry-characterized perturbations.}

\rev{Equation~\eqref{eq:Sndesc} describes the inverted Dirac sector,
not a globally regularized Bloch Hamiltonian. For $n\geq2$, the
transverse terms grow at least as fast as $Bk^2$; consequently, the
direction of $\bm d(\bk)$ at large $k$ depends on the polar angle.
The continuum momentum plane cannot then be compactified by assigning
a single limiting state at infinity. The cone sum in
Eq.~\eqref{eq:SCn} must therefore be regarded as the contribution of
this sector. A lattice regularization can contain additional band
inversions and modify the total Chern number. We establish the global
invariants below using exactly periodic Hamiltonians.}

\begin{figure}[t]
\centering
\includegraphics[width=0.96\textwidth]{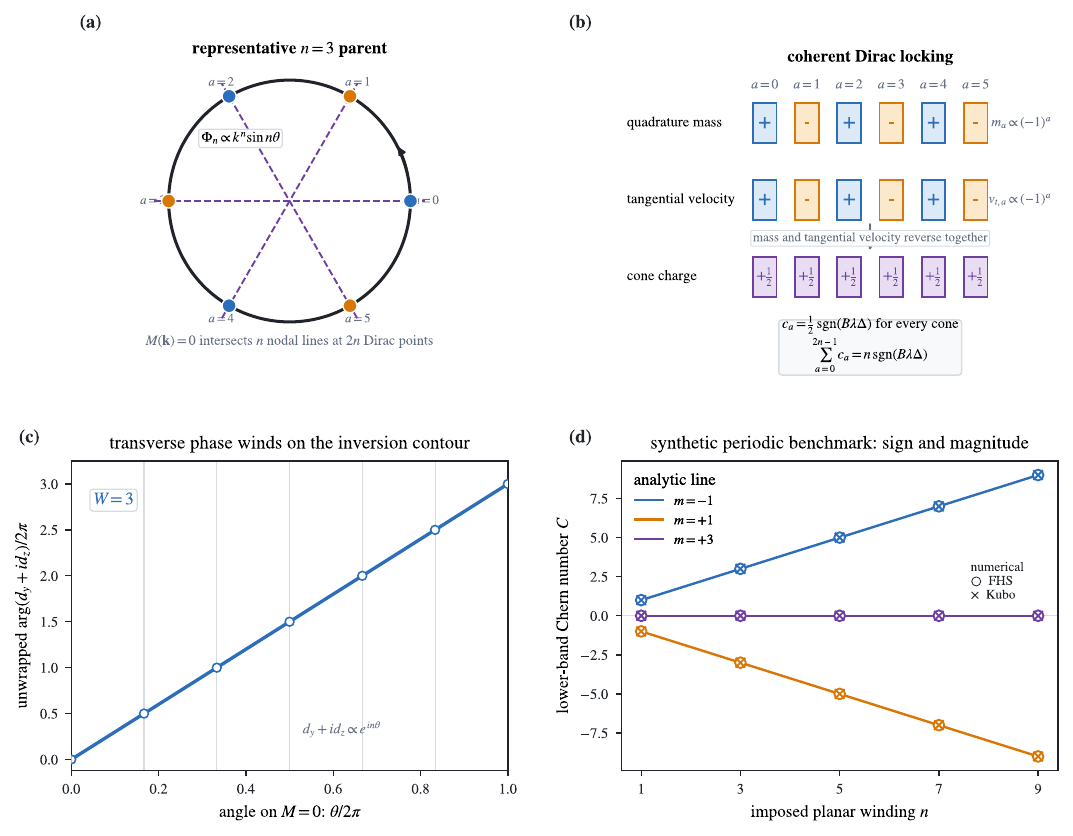}
\caption{\rev{\textbf{Planar winding criterion.} (a) For $n=3$, the inversion contour crosses three nodal lines at six Dirac points. (b) In the chosen quadrature, mass and tangential-velocity signs reverse together, giving equal half-Chern contributions. (c) The complex transverse pair winds three times. (d) Periodic tests with imposed windings $n=1,3,5,7,9$: analytic, FHS, and Kubo results agree. The high-$n$ entries are synthetic benchmarks, not assignments to crystallographic high-wave classes.}}
\label{fig:S1}
\end{figure}

\rev{Figure~\ref{fig:S1} illustrates the $n=3$ case and checks the periodic family}
\begin{align}
 H_n^{\square}(\bk)&=D_x\mu_x+D_y\mu_y+D_z\mu_z,\\
 D_x&=m+\cos k_x+\cos k_y,\\
 D_y+\ii D_z&=(\sin k_x+\ii\sin k_y)^n .
\label{eq:Ssquare}
\end{align}
\rev{For the odd-$n$ benchmarks, the lower-band invariant is}
\begin{equation}
 C=\frac{n}{2}\left[\sgn(m+2)+\sgn(m-2)-2\sgn m\right],
\label{eq:SsquareC}
\end{equation}
\rev{Thus $C=+n$ for $-2<m<0$, $C=-n$ for $0<m<2$, and $C=0$ outside
these intervals. FHS and direct Kubo integration reproduce these values.
The $n=5,7,9$ entries are synthetic planar benchmarks with imposed complex
winding. They are not the crystallographic $h$, $j$, and degree-nine
spin-splitting forms.}

\section{\texorpdfstring{Exact triangular-lattice $f$-wave regulator}{Exact triangular-lattice f-wave regulator}}
\label{sec:freg}

\rev{We next test the criterion in an exactly periodic triangular model.
Its primitive vectors are}
$\bm a_1=(1,0)$ and $\bm a_2=(1/2,\sqrt3/2)$ and
\begin{equation}
 \bm\delta_1=\bm a_1,\quad
 \bm\delta_2=\bm a_2-\bm a_1,\quad
 \bm\delta_3=-\bm a_2.
\end{equation}
With $\omega=e^{2\pi\ii/3}$ define
\begin{align}
 g(\bk)&=\sum_{j=0}^{2}\omega^j\sin(\bk\cdot\bm\delta_{j+1}),\label{eq:Sg}\\
 \Phi(\bk)&=\sum_{j=1}^{3}[1-\cos(\bk\cdot\bm\delta_j)].
\label{eq:Sgphi}
\end{align}
\rev{The harmonic obeys}
\begin{equation}
 g(C_3\bk)=\omega g(\bk),\qquad g(-\bk)=-g(\bk),\qquad
 g(M_y\bk)=g(\bk)^*
\label{eq:Sgsym}
\end{equation}
\rev{exactly. Near $\Gamma$, $g=(3/2)(k_x+\ii k_y)+O(k^3)$, so}
\begin{equation}
 \mathrm{Im}\,g^3=\frac{27}{8}k_y(3k_x^2-k_y^2)+O(k^5),
\end{equation}
\rev{which has the $f$-wave nodal form discussed in
Ref.~\cite{Hirschmann2026Fwave}. We specify the descendant as}
\begin{equation}
 H_f=[m-B\Phi]\mu_x+\lambda\mathrm{Re}(g^3)\mu_y
 +\Delta\mathrm{Im}(g^3)\mu_z .
\label{eq:Sfreg}
\end{equation}
\rev{This is a lattice realization of the angular mass pair, not a
material-specific microscopic Hamiltonian.}

\rev{Within one reciprocal primitive cell, $g$ vanishes at $\Gamma$, the three
$M$ points, and $K,K'$. Each has vorticity $+1$, $-1$, and $+1$,
respectively, with $\Phi=0,4,9/2$. Raising $g$ to an integer power
multiplies these vorticities by that power. In the next equation,
$\ell$ denotes this power only; it is not a universal assignment from a
crystallographic wave label. The lower-band Chern number is}
\begin{equation}
 C_\ell=\frac{\ell}{2}\sgn(\lambda\Delta)
 \sum_{\alpha\in Z_g}\chi_\alpha\sgn[m-B\Phi_\alpha].
\label{eq:Sglobalvort}
\end{equation}
For $\ell=3$,
\begin{equation}
 C_f=\frac32\sgn(\lambda\Delta)
 \left[\sgn m-3\sgn(m-4B)
 +2\sgn\!\left(m-\frac92B\right)\right].
\label{eq:Sfexact}
\end{equation}
For $B\lambda\Delta>0$, the phase sequence is
$0,+3,-6,0$, separated by $m/B=0,4,9/2$.

At the reference point $m/B=1.5$, $\lambda=1.2$, $\Delta=1$,
\begin{equation}
 C_f^{\rm Kubo}=C_f^{\rm FHS}=3.00,
\end{equation}
the minimum direct gap on the validation mesh is $2.739019B$, and the
unwrapped Wilson phase winds three times.  Setting the quadrature channel to
zero locates six Dirac nodes.  The local Jacobian and mass signs alternate,
but each charge is $+1/2$, exactly as in
Eq.~\eqref{eq:Shalf}.  At $m/B=4.25$ the analytic and FHS values are both
$C_f=-6$.

To test stability of the gapped $C=3$ descendant, we used a deterministic
96-sample scan (seed 260915), varying $m$ by $\pm24\%$, $B$ by $\pm18\%$,
and $\lambda,\Delta$ by $\pm35\%$.  We additionally added
\begin{equation}
 \delta d_y=\varepsilon[0.37\sin(\bk\cdot\bm\delta_1)
 +0.23\sin(\bk\cdot\bm\delta_2)],\qquad
 \varepsilon\in[-0.08,0.08],
\end{equation}
which explicitly breaks the organizing $C_3$ symmetry.  All 96 samples remain
at $C=3$ and the smallest sampled direct gap is $1.930B$.  \rev{This tests a finite neighborhood in Hamiltonian parameter space, not
spatial disorder. The gapped Chern phase survives the tested explicit
$C_3$ breaking; this does not establish which mass a material generates.}

\begin{figure}[t]
\centering
\includegraphics[width=0.96\textwidth]{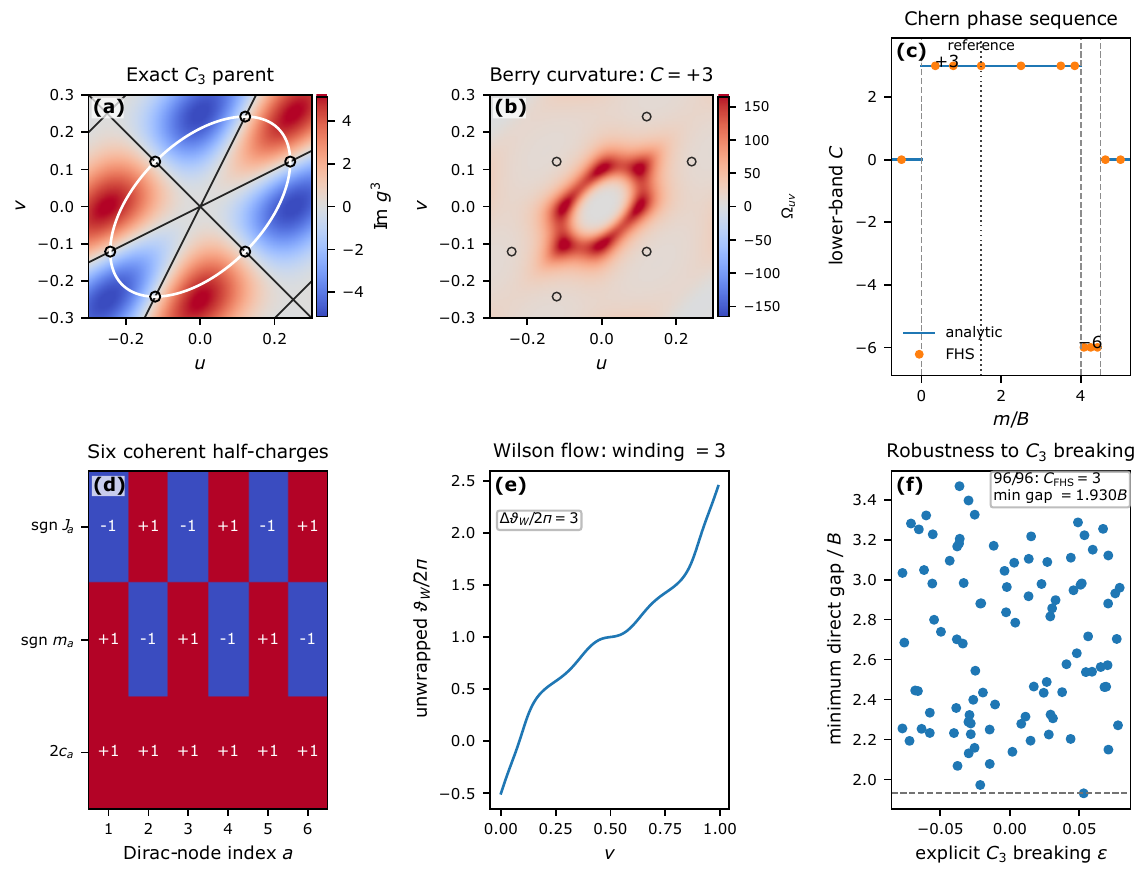}
\caption{\rev{\textbf{Triangular $f$-wave construction.} (a) Parent harmonic on the inversion contour. (b) Berry curvature for the specified $C=3$ descendant. (c) Chern sequence across the three lattice inversions. (d) Alternating Jacobian and mass signs give six equal half-charges. (e) Wilson-loop winding is three. (f) All 96 tested parameter sets, including explicit $C_3$ breaking, retain $C=3$; the smallest sampled gap is $1.930B$.}}
\label{fig:S2}
\end{figure}

\section{\texorpdfstring{Crystallographic $h$ and $j$ forms with specified mass partners}{Crystallographic h and j forms with specified mass partners}}
\label{sec:highwaves}

\rev{For coplanar tetragonal and hexagonal spin point groups, the
classification of Ref.~\cite{Elcoro2026HighOrder} gives the real
odd-parity basis functions}
\begin{align}
 \Phi_h(\bk)&\propto k_xk_yk_z(k_x^2-k_y^2)
 =\frac{k_z}{4}\operatorname{Im}(k_x+\ii k_y)^4,
 \label{eq:Shpoly}\\
 \Phi_j(\bk)&\propto k_xk_yk_z(3k_x^2-k_y^2)(k_x^2-3k_y^2)
 =\frac{k_z}{2}\operatorname{Im}(k_x+\ii k_y)^6.
 \label{eq:Sjpoly}
\end{align}
\rev{These are degree-five and degree-seven functions, but their
in-plane angular orders at fixed $k_z\ne0$ are four and six.
The forms specify an odd-parity parent splitting, not a complex mass
winding. In the constructions below, the inversion term and the
real cosine quadrature are \emph{additional model ingredients}.
They respect the organizing rotations, but are not asserted to be
mandatory SOC or exchange terms of every material in the corresponding
spin-point-group class.}

\begin{figure}[t]
\centering
\includegraphics[width=0.98\textwidth]{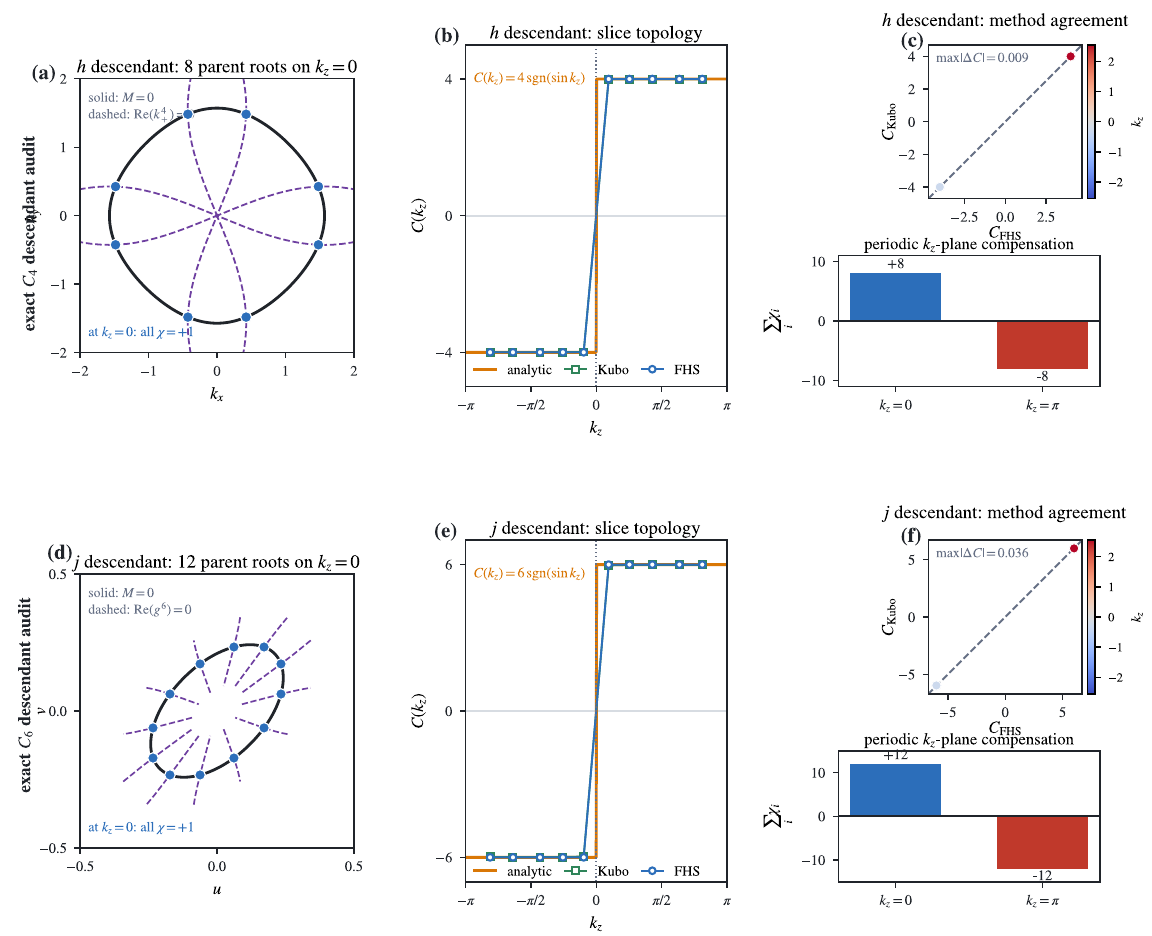}
\caption{\rev{\textbf{Topology of the specified $h$- and $j$-wave descendants.}
(a,d) The equations $d_x=d_y=0$ give eight $h$-wave and twelve $j$-wave
Weyl points on $k_z=0$. The plotted positive couplings give charge
$+1$ at each point; $(u,v)$ are triangular reciprocal coordinates in
(d). (b,e) The analytic slice Chern numbers are confirmed by FHS and
Kubo integration. They are defined only for gapped $k_z$ slices, not
on the nodal planes. (c,f) Numerical comparison and total Weyl charges
$+8,-8$ and $+12,-12$ on $k_z=0,\pi$ respectively, demonstrating
periodic charge compensation.}}
\label{fig:S3}
\end{figure}

\rev{For the tetragonal lattice define
$z(\bk_\parallel)=\sin k_x+\ii\sin k_y$. It satisfies}
\begin{equation}
 z(C_4\bk_\parallel)=\ii z(\bk_\parallel),\quad
 z(-\bk_\parallel)=-z(\bk_\parallel),\quad
 z(M_y\bk_\parallel)=z(\bk_\parallel)^*.
\label{eq:Shsymmetry}
\end{equation}
\rev{Thus $z^4$ is fourfold invariant and even under in-plane momentum
inversion; its real and imaginary parts have opposite parities under
the vertical reflection $M_y:(k_x,k_y)\mapsto(k_x,-k_y)$.
A fully periodic model with the specified pair is}
\begin{equation}
 H_h=[m+\cos k_x+\cos k_y]\mu_x
 +\lambda\operatorname{Re}(z^4)\mu_y
 +\Delta\sin k_z\operatorname{Im}(z^4)\mu_z.
\label{eq:Shreg}
\end{equation}
\rev{The small-momentum form of its last term is proportional to
$k_z\operatorname{Im}(k_x+\ii k_y)^4$, as required of the chosen
parent basis function. For $-2<m<0$ and $\lambda\Delta\sin k_z\ne0$,
there are no zeros of the three-component $\bm d$ vector on a fixed
$k_z$ slice: $z=0$ only at $k_x,k_y=0,\pi$, where the possible
inversion masses are $m+2$, $m$, and $m-2$, none of which vanishes in
this interval. Its lower-band Chern number is}
\begin{equation}
 C_h(k_z)=4\,\sgn[\lambda\Delta\sin k_z].
\label{eq:Shslice}
\end{equation}
\rev{On the plane $k_z=0$, $d_z$ vanishes identically. The surviving
crossings consequently solve $d_x=d_y=0$, i.e., the inversion contour
intersects $\operatorname{Re}(z^4)=0$ at eight points. These are not to
be confused with the zeros of the parent sine harmonic on a generic
nonzero-$k_z$ slice. The linearized Jacobian
$\det[\partial(d_x,d_y,d_z)/\partial(k_x,k_y,k_z)]$ has the same sign
at all eight points. For the positive couplings used in Fig.~\ref{fig:S3},
the lower-band Weyl charge is $+1$ per node, and the Chern number jumps
by $+8$ across $k_z=0$. At $k_z=\pi$ the derivative of $\sin k_z$
reverses, producing eight $-1$ nodes and total charge $-8$.}

\rev{For the hexagonal construction, use the three-bond triangular
harmonic $g$ and invariant $\Phi$ from
Eqs.~\eqref{eq:Sg} and \eqref{eq:Sgphi}. In addition to
Eq.~\eqref{eq:Sgsym}, they satisfy}
\begin{equation}
 g(C_6\bk_\parallel)=e^{\ii\pi/3}g(\bk_\parallel),\qquad
 \Phi(C_6\bk_\parallel)=\Phi(\bk_\parallel).
\label{eq:Sjsymmetry}
\end{equation}
\rev{Consequently $g^6$ and $\Phi$ are invariant under the lattice
60-degree momentum rotation, and $g^6$ is even under in-plane
inversion. The exact lattice descendant is}
\begin{equation}
 H_j=[m-B\Phi(\bk_\parallel)]\mu_x
 +\lambda\operatorname{Re}(g^6)\mu_y
 +\Delta\sin k_z\operatorname{Im}(g^6)\mu_z.
\label{eq:Sjreg}
\end{equation}
\rev{Near $\Gamma$, $g=(3/2)(k_x+\ii k_y)+O(k^3)$, so the last term has
the degree-seven form $k_z\operatorname{Im}(k_x+\ii k_y)^6$ to leading
order. In the first inverted regime $0<m/B<4$, with
$\lambda\Delta\sin k_z\ne0$, the gapped slices carry}
\begin{equation}
 C_j(k_z)=6\,\sgn[B\lambda\Delta\sin k_z].
\label{eq:Sjslice}
\end{equation}
\rev{Here $g$ can vanish only at $\Gamma$, the three $M$ points,
and $K,K'$, where $\Phi=0$, $4$, and $9/2$, respectively
(Sec.~\ref{sec:freg}); none lies on $m-B\Phi=0$ in this regime.
At $k_z=0$, $d_x=d_y=0$ has twelve isolated solutions in a
reciprocal primitive cell. Their Jacobians have the same sign;
for the plotted couplings each has lower-band Weyl charge $+1$.
Twelve compensating $-1$ nodes occur at $k_z=\pi$. The two node planes
therefore contribute $+12$ and $-12$ respectively, as required by
charge neutrality in a periodic Brillouin zone~\cite{Wan2011Weyl,Armitage2018Weyl}.}

\rev{Both periodic models possess a useful \emph{effective} antiunitary
relation. Because $d_x$ and $d_y$ are even under full momentum
inversion while $d_z$ is odd, $\Theta=\mu_x K$ obeys}
\begin{equation}
 \Theta H_{h,j}(\bk)\Theta^{-1}=H_{h,j}(-\bk),\qquad
 \Theta^2=+1.
\label{eq:Shjantiunitary}
\end{equation}
\rev{This is a property of the constructed pseudospin Hamiltonians, not
an identification with microscopic time reversal of a particular
magnetic order. It relates opposite momentum slices and imposes
$C_{h,j}(k_z)=-C_{h,j}(-k_z)$. It does not preserve a generic
nonzero-$k_z$ slice and therefore does not prohibit its nonzero
Chern number. The planes $k_z=0,\pi$ are gapless, so no slice invariant
is assigned there. Likewise, the $M_y$ parity of the scalar functions
above is not by itself a full mirror symmetry of a magnetic material:
that would also require the action of the mirror on the doublet.}

\rev{We evaluate the slice invariants using the periodic two-band
Hamiltonians, checking the winding prediction against Fukui--Hatsugai--Suzuki
(FHS) lattice-gauge sums and direct $\hat{\bm d}$-curvature integration.
The representative finite-mesh discrepancies are of order $10^{-2}$,
while the FHS values and analytic charges give the plateaus
$\pm4$ and $\pm6$ (Fig.~\ref{fig:S3}). Independent solutions of
$d_x=d_y=0$ and Jacobian signs give the node counts and chiralities.
These tests establish the stated \emph{model} topology, not the
symmetry permission or microscopic strength of the assumed quadrature
in any particular material.}

\rev{We do not extend the same angular reduction to the degree-nine
class. Its allowed three-dimensional polynomial does not provide a
single longitudinal factor times one planar harmonic with a unique
mass winding. Material-specific conclusions would require the
representations of an isolated doublet and all permitted transverse
terms, not the wave label alone.}

\section{\texorpdfstring{\rev{Microscopic four-band $p$-wave model and mass constraint}}{Microscopic four-band p-wave model and mass constraint}}
\label{sec:pwave}

\subsection{Parent, symmetries, and orbital sewing}

\rev{The parent model acts on two site ($\tau$) and spin ($\sigma$)
degrees of freedom and is adapted from the established minimal
$p$-wave Hamiltonian~\cite{Hellenes2023Pwave,Chakraborty2025Edelstein}:}
\begin{align}
H_p(\bk)=&\;2t\cos\frac{k_x}{2}\tau_x+2t_y\cos k_y\,\id
+2J\sin\frac{k_x}{2}\tau_y\sigma_x\notag\\
&+[\delta_0+2J\cos k_y]\tau_z\sigma_y .
\label{eq:Sparent}
\end{align}
\rev{The canonical model is recovered at $t_y=t$ and $\delta_0=0$.
The $t_y$ deformation and the additional $\delta_0\tau_z\sigma_y$
term preserve the reduced symmetry set below; $\delta_0$ removes an
accidental secondary zero in the subsequent nodal-line analysis.
Neither modification is identified with a particular experimental field.}

\rev{In the symmetric orbital gauge, define the two antiunitaries and
the unitary $x$-reversing operation}
\begin{equation}
 \Xi=\ii\sigma_xK,\qquad
 \Tt=\tau_x\ii\sigma_yK,\qquad
 U_y=\ii\sigma_y,
\label{eq:Sops}
\end{equation}
\rev{with $K$ denoting complex conjugation. Their actions are}
\begin{align}
 \Xi H_p(\bk)\Xi^{-1}&=H_p(-\bk),\\
 \Tt H_p(\bk)\Tt^{-1}&=H_p(-\bk),\\
 U_yH_p(k_x,k_y)U_y^{-1}&=H_p(-k_x,k_y).
\end{align}
\rev{In this representation, $\Xi^2=+1$ and $\Tt^2=-1$.
The antiunitaries are effective symmetries of the Bloch Hamiltonian;
they are not equated with microscopic time reversal of the ordered
magnetic texture~\cite{Luo2026Hidden}. Their product is, up to phase,
the fixed-momentum unitary symmetry
$\mathcal S=\tau_x\sigma_z$. It commutes with $H_p(\bk)$ and
anticommutes with $U_y$. Since $U_y$ maps $k_x=0$ to itself,
$\{\mathcal S,U_y\}=0$ forces states of opposite $\mathcal S$
eigenvalues to be degenerate on that line. This gives two doublets,
which may be split away from $k_x=0$.}

\rev{The symmetric two-site Bloch basis is not strictly periodic
under $k_x\to k_x+2\pi$; instead,}
\begin{equation}
 H_p(k_x+2\pi,k_y)=G_xH_p(k_x,k_y)G_x^\dagger,
 \qquad G_x=\tau_z .
\label{eq:Ssew}
\end{equation}
\rev{Both descendant terms below obey the same sewing relation.
The $G_x$ transformation must therefore be included in FHS and
Wilson overlaps crossing the $k_x$ boundary.}

\rev{Introduce}
\begin{equation}
 a=2t\cos\frac{k_x}{2},\quad b=2J\sin\frac{k_x}{2},\quad
 d=\delta_0+2J\cos k_y,\quad c=2t_y\cos k_y.
\end{equation}
\rev{The exact parent energies are the sorted set}
\begin{equation}
 E_{u,\pm}=c\pm\sqrt{a^2+(b-u d)^2},\qquad u=\pm1.
\label{eq:Sspectrum}
\end{equation}
\rev{Direct numerical diagonalization reproduces this expression.}

\subsection{Nodal-doublet projection and complete mass basis}

\rev{On the nodal line $k_x=0$, let $a=2t$,
$d(k_y)=\delta_0+2J\cos k_y$, and $R=\sqrt{a^2+d^2}$.
The doublet energies are $\epsilon_s=c+sR$, $s=\pm1$, with
projectors}
\begin{equation}
 P_s=\frac12\left[\id+s\frac{a\tau_x+d\tau_z\sigma_y}{R}\right].
\label{eq:Sproj}
\end{equation}
\rev{Within either projected doublet, a convenient Pauli basis is}
\begin{equation}
 \mu_x=\tau_x\sigma_x,\qquad
 \mu_y=\sigma_y,\qquad
 \mu_z=\tau_x\sigma_z.
\label{eq:Smu}
\end{equation}
\rev{To leading order in the normal momentum $q_x=k_x$,}
\begin{equation}
 h_s=\epsilon_s(k_y)+v_s(k_y)q_x\mu_z+O(q_x^2),
 \qquad v_s=-sJ\frac{d(k_y)}{R(k_y)}.
\label{eq:Svel}
\end{equation}
\rev{For $\delta_0>2|J|$, the factor $d(k_y)$ never changes sign;
with positive $J$, the lower-doublet velocity $v_-$ is positive
throughout the line.}

\rev{Table~\ref{tab:Sall16} gives the complete first-order projection
of the 16 Hermitian matrices $\tau_a\sigma_b$.
Here ``first order'' refers to the perturbation amplitude; the
projection is evaluated at $q_x=0$ for generic $k_y$.
Six matrices have nonzero components transverse to $\mu_z$ and can
open a gap in degenerate perturbation theory. The remaining ten have
scalar, longitudinal, or zero projections. A null projection at this
order does not exclude effects away from the line or at higher order.}

\begin{table}[t]
\caption{Projection of all 16 Hermitian internal matrices onto the $s$
nodal doublet. The characters $\pm$ refer to the internal matrix
before the momentum dependence of its coefficient is included.
Here $\alpha=a/R$ and $\beta=d/R$; $L$ denotes longitudinal,
$S$ scalar, and $0$ null projections.}
\label{tab:Sall16}
\begin{ruledtabular}
\begin{tabular}{lcccccl}
Matrix & $\Xi$ & $\Tt$ & $U_y$ & $\mathcal S$ & Projection & Role\\
\hline
$\id$                   & + & + & + & + & $\id$ & scalar\\
$\sigma_x$              & + & - & - & - & $s\alpha\mu_x$ & mass\\
$\sigma_y$              & + & - & + & - & $\mu_y$ & mass\\
$\sigma_z$              & - & - & - & + & $s\alpha\mu_z$ & longitudinal\\
$\tau_x$                & + & + & + & + & $s\alpha\id$ & scalar\\
$\tau_x\sigma_x$        & + & - & - & - & $\mu_x$ & mass\\
$\tau_x\sigma_y$        & + & - & + & - & $s\alpha\mu_y$ & mass\\
$\tau_x\sigma_z$        & - & - & - & + & $\mu_z$ & longitudinal\\
$\tau_y$                & - & + & + & - & $0$ & null\\
$\tau_y\sigma_x$        & - & - & - & + & $-s\beta\mu_z$ & longitudinal\\
$\tau_y\sigma_y$        & - & - & + & + & $0$ & null\\
$\tau_y\sigma_z$        & + & - & - & - & $s\beta\mu_x$ & mass\\
$\tau_z$                & + & - & + & - & $s\beta\mu_y$ & mass\\
$\tau_z\sigma_x$        & + & + & - & + & $0$ & null\\
$\tau_z\sigma_y$        & + & + & + & + & $s\beta\id$ & scalar\\
$\tau_z\sigma_z$        & - & + & - & - & $0$ & null\\
\end{tabular}
\end{ruledtabular}
\end{table}

\rev{For each of the six transverse mass matrices $M$, direct matrix
conjugation gives}
\begin{equation}
 \Xi M\Xi^{-1}=+M,\qquad
 \Tt M\Tt^{-1}=-M,\qquad
 \mathcal S M\mathcal S^{-1}=-M .
\label{eq:Smasschars}
\end{equation}
\rev{A real term $f(\bk)M$ thus respects $\Xi$ only if
$f(-\bk)=f(\bk)$ and respects $\Tt$ only if
$f(-\bk)=-f(\bk)$. No nonzero first-order transverse
mass can preserve both antiunitaries. This exhausts the internal
four-band basis at the parent line; it is not a prohibition on
higher-order effects or a classification of every $p$-wave magnet.
The two transverse families are also distinct under $U_y$:
$\mu_x$ masses are $U_y$-odd, while $\mu_y$ masses are $U_y$-even.}

\subsection{Complex mass winding and four-band topology}

\rev{Near the lower or upper parent nodal doublet, two complementary
projected masses give}
\begin{equation}
 h_s(q_x,k_y)=\epsilon_s(k_y)+v_s(k_y)q_x\mu_z
 +m_1(k_y)\mu_x+m_2(k_y)\mu_y .
\label{eq:Seff}
\end{equation}
\rev{Write $m_1+\ii m_2=|m|e^{\ii\varphi}$. With the Berry-curvature
convention of Sec.~I, the curvature of the lower eigenstate of this
linearized two-band Hamiltonian is}
\begin{equation}
 \Omega_-^z(q_x,k_y)=
 -\frac{v_s |m|^2\partial_{k_y}\varphi}
 {2\left(|m|^2+v_s^2q_x^2\right)^{3/2}} .
\label{eq:Slineberry}
\end{equation}
\rev{This expression is exact within Eq.~\eqref{eq:Seff}; terms
beyond the linearized description need not share its simple form.
Integrating over the normal direction yields the nodal-line contribution}
\begin{equation}
 C_{\rm line}=-\sgn(v_s)\frac{1}{2\pi}\oint d\varphi
 =-\sgn(v_s)W_s .
\label{eq:SlineC}
\end{equation}
\rev{This formula requires a nonvanishing mass along the line and a
normal velocity of fixed sign. It describes an isolated sector; the
periodic four-band Chern invariant is evaluated separately.
It is not applied directly to the intersecting higher-wave
nodal networks.}

\rev{For an explicit test, choose the momentum-odd spin coupling and
the momentum-even exchange-like channel}
\begin{align}
 H_{\rm r}(\bk)&=\lambda\sin k_y\,\sigma_y,\\
 H_{\rm b}(\bk)&=[M_0+M_x\cos k_x+M_y\cos k_y]\sigma_x .
\label{eq:Sdesc}
\end{align}
\rev{The $H_{\rm r}$ term breaks $\Xi$ while preserving $\Tt$ and
$U_y$; $H_{\rm b}$ preserves $\Xi$ but breaks $\Tt$ and $U_y$.
Their combination breaks all three symmetries without changing
orbital sewing. Either surviving antiunitary would prohibit a
nonzero first Chern number for any isolated band in two dimensions;
so would the orientation-reversing $U_y$ if it remained unbroken.
An individual perturbation also need not open a global gap. Under
bare time reversal $\mathcal T=\ii\sigma_yK$, $H_{\rm r}$ is even
and $H_{\rm b}$ is odd, while the magnetic parent itself breaks
$\mathcal T$. Thus the two channels have SOC-like and exchange-like
symmetry characters, respectively. They have not been microscopically
assigned to a specific SOC or exchange perturbation in a material.}

\rev{For the lower parent doublet, the transverse components are}
\begin{align}
 m_1(k_y)&=-\frac{a}{R(k_y)}[M_0+M_x+M_y\cos k_y],\\
 m_2(k_y)&=\lambda\sin k_y .
\label{eq:Smassloop}
\end{align}
\rev{At $M_x=0$, positive $t$, and $\lambda M_y>0$, the complex
mass winds $W_-=-1$ for $|M_0|<|M_y|$ and winds zero times
for $|M_0|>|M_y|$. The gap closes at $M_0/M_y=\pm1$; for nonzero $M_x$,
the nodal-line condition shifts to $M_0+M_x=\pm M_y$.
Both regimes use the same symmetry-characterized mass channels;
therefore, the allowed operators alone do not determine the topology.}

\begin{figure}[t]
\centering
\includegraphics[width=0.98\textwidth]{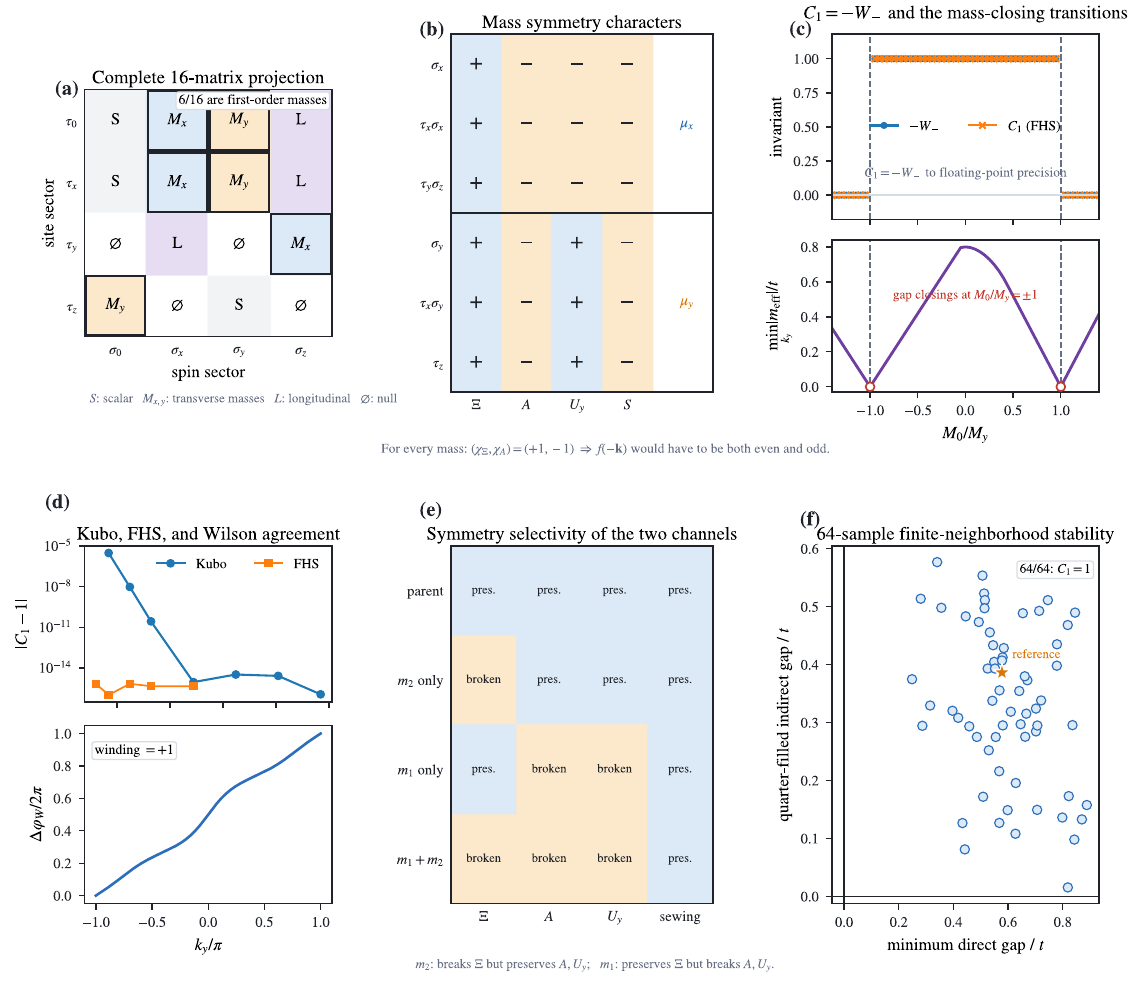}
\caption{\rev{\textbf{Microscopic $p$-wave mass analysis.}
(a) Complete projection of 16 internal matrices: six are first-order
transverse masses.
(b) Every transverse mass is even under $\Xi$, odd under $\Tt$,
and odd under their fixed-momentum product $\mathcal S$.
(c) Lowest-band Chern number and the opposite projected mass winding
across the two mass closures $M_0/M_y=\pm1$.
(d) Kubo and sewing-aware FHS agreement, together with the
$+1$ Wilson-loop winding for the topological reference state.
(e) Symmetry actions of the two perturbation channels; orbital
boundary sewing remains intact.
(f) All 64 sampled parameter variations retain the lowest-band
$C_1=1$ phase and a quarter-filled gap.}}
\label{fig:S4}
\end{figure}

\rev{We choose the reference parameters}
\begin{equation}
 t=1,\ J=0.25,\ \delta_0=1.2,\ t_y=0.25,\ \lambda=0.8,
 \ M_0=-0.15,\ M_x=0,\ M_y=1.1 .
\label{eq:Sparams}
\end{equation}
\rev{The adjacent direct gaps are approximately $(1.00,0.58,1.00)$
and the indirect gap at quarter filling is $0.39$ in units of $t$.
With the orbital boundary sewing included, a $41\times41$ FHS
calculation gives}
\begin{equation}
 (C_1,C_2,C_3,C_4)=(+1,-1,+1,-1).
\end{equation}
\rev{Thus the isolated lowest band has $C_1=1$, while the first two
bands have zero total Chern number. The lower-band Chern invariant
agrees with direct Kubo integration and the net Wilson-loop winding:}
\begin{equation}
 C_1^{\rm Kubo}=C_1^{\rm FHS}=W_{\rm Wilson}=1
\quad\text{(within numerical precision)}.
\end{equation}
\rev{The Wilson phase winds in the opposite direction to the projected
mass phase because $C_{\rm line}=-\sgn(v_-)W_-$ and $v_->0$.
Figure~\ref{fig:S4}(c) traces the winding change across the two
mass closures, while panels (d,e) check the invariant and symmetries.}

\rev{Finally, a deterministic finite-neighborhood test varied
$t$ by $\pm10\%$, $J$ by $\pm8\%$, $\delta_0$ by $\pm8\%$,
$t_y$ by $\pm20\%$, $\lambda$ by $\pm15\%$,
$M_0$ by $\pm30\%$, $M_y$ by $\pm10\%$,
and $M_x\in[-0.05,0.05]$.
All 64 tested parameter sets retained $C_1=1$, a direct gap,
and a positive quarter-filled indirect gap [Fig.~\ref{fig:S4}(f)].
The smallest sampled direct and indirect gaps were about
$0.25t$ and $0.015t$, respectively. This is a finite-parameter
stability test of the model, not a disorder calculation or a
material-specific prediction.}

\section{First-principles protocol and material context}
\label{sec:dftmethods}

\rev{Our reference system is the metallic LaMn$_{2-x}$Au$_{4+x}$ compound
at $x\simeq1$. Powder-neutron diffraction is compatible with a coplanar
hedgehog spin-vortex crystal (SVC) and a competing collinear arrangement.
The SVC refinement has magnetic space group $I_C4cm$; relativistic
first-principles calculations reported for the compound favor this
state~\cite{Siebeneichler2024LaMnAu5}. We use the reported SVC to
examine the symmetry-dependent electronic structure. Its $h$-wave
spin-splitting character was identified by the spin-point-group and
anti-spin-Laue classifications~\cite{Elcoro2026HighOrder,Lange2026ASLG}.}

\rev{The electronic calculations use the projector-augmented-wave
implementation of density-functional theory in VASP with the
Perdew--Burke--Ernzerhof functional
\cite{Kresse1996VASP,Kresse1999PAW,Perdew1996PBE}. The reported production
settings include a plane-wave cutoff of $550$~eV, a reciprocal-mesh
density of $2\pi\times0.04\,\mathrm{\AA}^{-1}$, and an electronic
energy-convergence threshold of $10^{-6}$~eV. Magnetism is treated
noncollinearly, without imposing spatial symmetries in the electronic
minimization. The relativistic comparisons use the reference SVC and
two configurations with all Mn moments coherently rotated in the plane
by $\phi=\pm22.5^\circ$. Results are compared on matching momentum
paths, with energies referenced to the corresponding calculation's
Fermi level where appropriate.}

\rev{To characterize the relevant low-energy states we complement the
sorted band indices with orbital/projector character, continuity of
the band-pair midpoint, and spin character. The spin information is
particularly useful without SOC, whereas projector and subspace
continuity carry greater weight with SOC. The separate character
tracking identifies band reorderings that cannot be resolved by
selecting the smallest gap between adjacent sorted eigenvalues.
}

\section{Nonrelativistic LaMnAu$_5$: the $h$-wave nodal fingerprint}
\label{sec:nosoc}

\rev{We first examine the positive energy separation of the targeted
low-energy bands on a $15\times15$ grid at fixed reduced $k_z=0.10$,
using}
\begin{equation}
\Delta(\bk)=|E_{634}(\bk)-E_{633}(\bk)|.
\end{equation}
\rev{Because this is a magnitude, it cannot on its own establish the
sign of the odd-parity splitting. The $h$-wave identification relies on
the classified magnetic symmetry, the four nodal directions, the
angular dependence of the spectrum, and independent orbital/spin
tracking. At fixed nonzero $k_z$, the leading positive $h$-wave form is}
\begin{equation}
\Delta_h^{(0)}(\bk)=A\left|k_z k_xk_y(k_x^2-k_y^2)\right|.
\label{eq:Sdfth}
\end{equation}
\rev{On the central domain $|k_x|,|k_y|\leq0.12$, excluding zeros of the
leading basis, this form gives $R^2\simeq0.997$ and a root-mean-square
fit error of about $0.16$~meV. Adding a same-symmetry radial correction,}
\begin{equation}
\Delta_h^{(2)}=|k_z k_xk_y(k_x^2-k_y^2)|
 [A+B(k_x^2+k_y^2)],
\label{eq:Sdftradial}
\end{equation}
\rev{gives $R^2\simeq0.999$ and an error near $0.05$~meV. This correction
does not alter the nodal directions. Here $A$ and $B$ are fit
parameters, unrelated to the coefficients of the lattice Hamiltonians.}

\rev{For a set of sampled momenta $S$, define the RMS splitting}
\begin{equation}
\Delta_{\mathrm{RMS}}(S)=
 \left[\frac{1}{N_S}\sum_{\bk\in S}\Delta(\bk)^2\right]^{1/2}.
\end{equation}
\rev{The resulting values over the full grid are}
\begin{center}
\begin{tabular}{lc}
\toprule
momentum sector & $\Delta_{\mathrm{RMS}}$ (meV)\\
\midrule
$k_x=0$ & 0.070\\
$k_y=0$ & 0.072\\
$k_x=k_y$ & 0.13\\
$k_x=-k_y$ & 0.17\\
all four nodal directions & 0.12\\
generic points & 13.48\\
\bottomrule
\end{tabular}
\end{center}
\rev{The combined nodal splitting is therefore less than one percent of
the generic splitting. The nodal suppression follows the symmetry-allowed
angular structure rather than an isotropic small-gap region.}

\begin{figure}[t]
\centering
\includegraphics[width=0.98\textwidth]{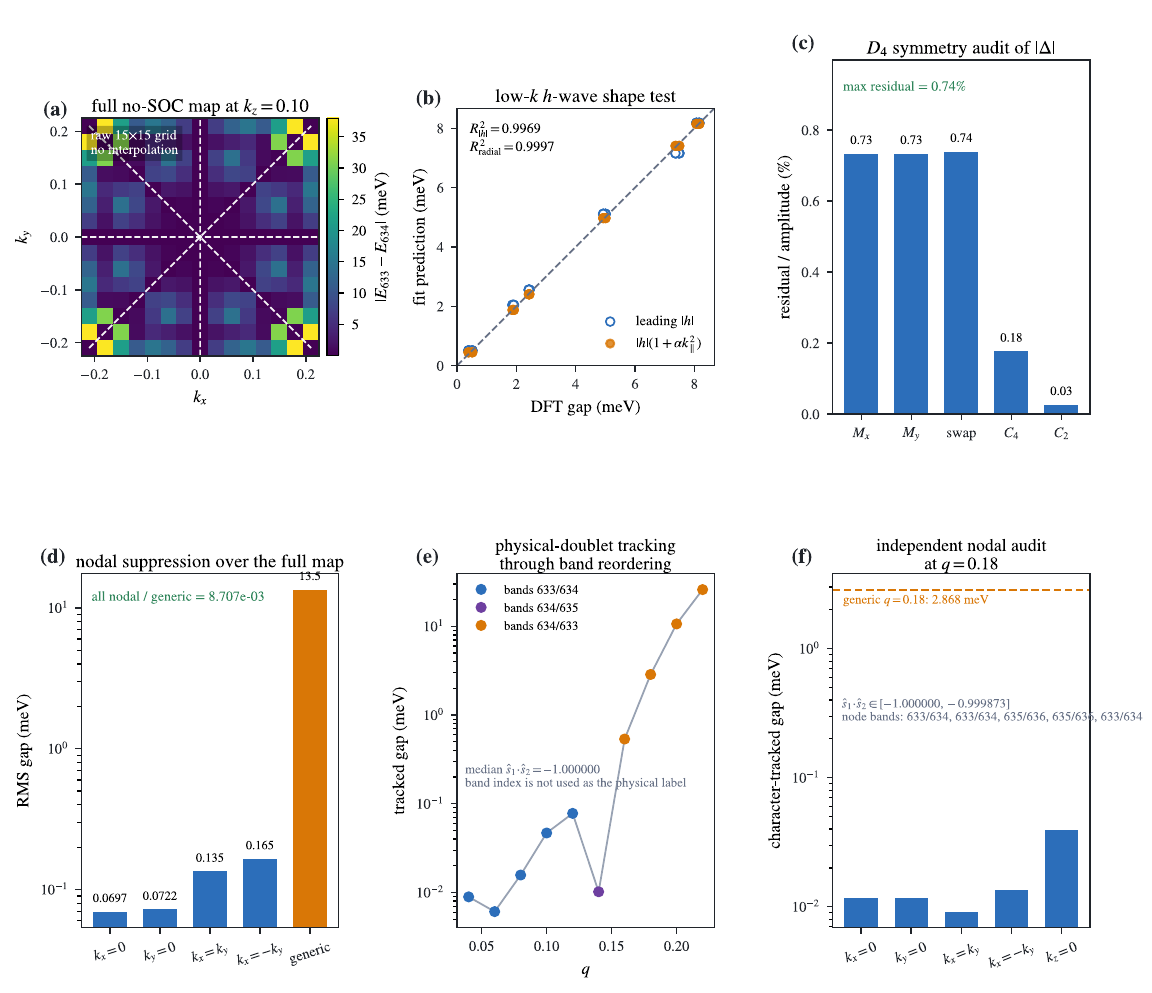}
\caption{\rev{\textbf{Nonrelativistic $h$-wave fingerprint in LaMnAu$_5$.}
(a) Directly sampled low-energy splitting at fixed $k_z=0.10$.
(b) Leading $h$-wave fit and same-symmetry radial correction.
(c) Residuals under the indicated tetragonal operations, evaluated for the
positive splitting magnitude. (d) Suppressed splitting on the four
parent nodal directions compared with generic momenta.
(e) Character-based band tracking across neighboring-band reorderings.
(f) Independent nodal-direction comparison along a second momentum path.
Reported quantities are evaluated on sampled data, without fitting to
interpolated display values.}}
\label{fig:S5}
\end{figure}

\rev{To quantify the symmetry of the splitting \emph{magnitude}, we use}
\begin{equation}
r_g=\frac{\sqrt{\langle[\Delta(\bk)-\Delta(g\bk)]^2\rangle}}
 {\sqrt{\left\langle[(\Delta(\bk)+\Delta(g\bk))/2]^2\right\rangle}}.
\label{eq:Ssymres}
\end{equation}
\rev{On the sampled grid the residuals under $M_x$, $M_y$, coordinate
exchange, $C_4$, and $C_2$ are about $0.73\%$, $0.73\%$, $0.74\%$,
$0.18\%$, and $0.03\%$. These values characterize the positive-gap
pattern, not the full signed spin texture. Along a separate generic path,
character tracking follows the physical doublet through changes of
its sorted VASP band labels. Independent nodal-direction profiles show
that the small splitting is retained while a generic direction acquires
a substantially larger separation [Fig.~\ref{fig:S5}].}

\section{Relativistic reconstruction of the $h$-wave parent}
\label{sec:soc}

\rev{The comparison with SOC uses the central $9\times9$ momentum grid.
The fixed-index doublet and independently selected adjacent pairs are
checked by their midpoint relative to $E_F$. The relevant $633/634$ pair
is identified throughout this central SOC comparison window.}

\begin{figure}[t]
\centering
\includegraphics[width=0.98\textwidth]{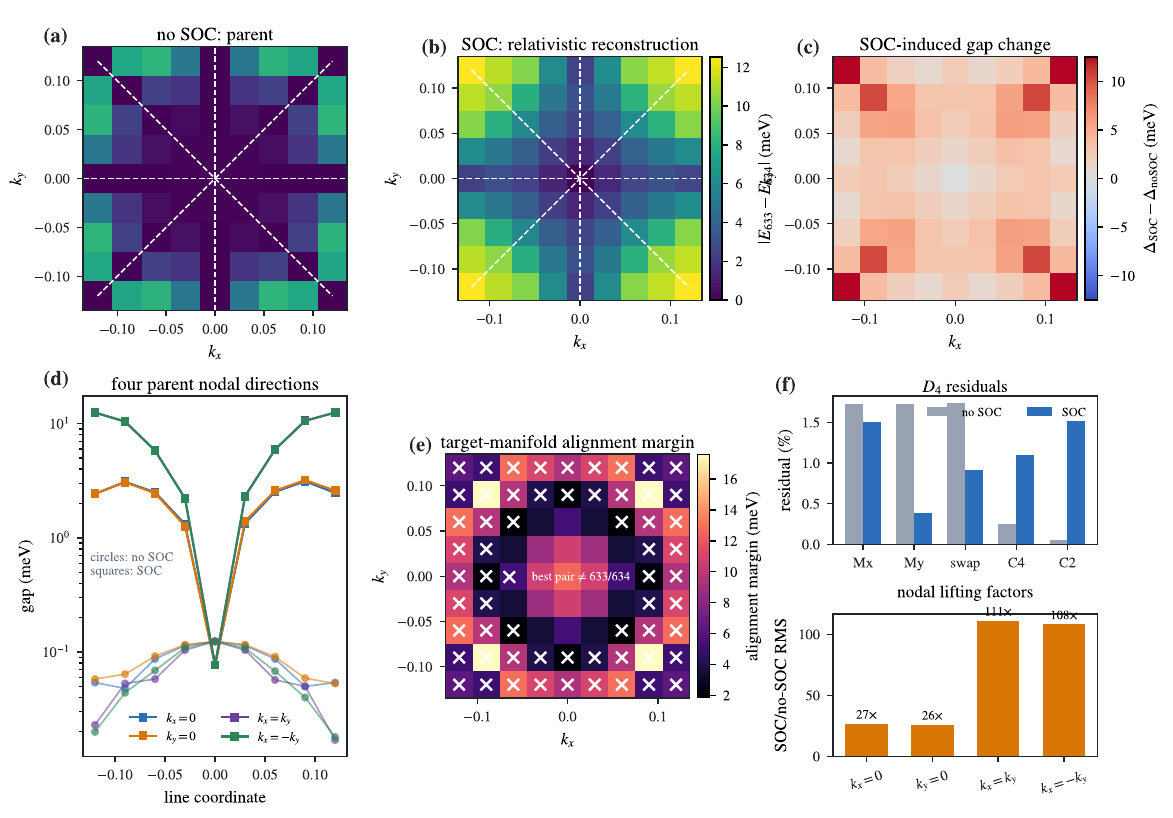}
\caption{\rev{\textbf{Relativistic reconstruction of the $h$-wave parent.}
(a,b) Directly sampled no-SOC and SOC doublet splittings on a common
momentum grid and energy scale. (c) Their pointwise difference.
(d) Splitting profiles on the four parent nodal directions.
(e) Independent alignment test for the target SOC doublet.
(f) Sampled symmetry-residual diagnostics and the enhancement of splitting
along the parent nodal directions. The maps show positive energy
separations rather than signed transverse masses.}}
\label{fig:S6}
\end{figure}

\rev{The RMS splittings on the four parent nodal directions are}
\begin{center}
\begin{tabular}{lccc}
\toprule
line & no SOC (meV) & SOC (meV) & enhancement\\
\midrule
$k_x=0$ & 0.086 & 2.30 & $27\times$\\
$k_y=0$ & 0.090 & 2.33 & $26\times$\\
$k_x=k_y$ & 0.074 & 8.26 & $111\times$\\
$k_x=-k_y$ & 0.076 & 8.26 & $108\times$\\
\bottomrule
\end{tabular}
\end{center}
\rev{The no-SOC values in this table are from the central comparison
grid, rather than the larger grid used above. SOC produces a much larger
splitting on the diagonal directions than on the axial directions.
The corresponding pointwise change in positive splitting,
$\Delta_{\mathrm{SOC}}(\bk)-\Delta_{\mathrm{noSOC}}(\bk)$, and the directly
sampled directional profiles appear in Fig.~\ref{fig:S6}.}

\rev{The symmetry-residual measure in Eq.~\eqref{eq:Ssymres} gives about
$(1.5,0.4,0.9,1.1,1.5)\%$ for
$(M_x,M_y,k_x\leftrightarrow k_y,C_4,C_2)$ with SOC.
These are tests of the sampled positive splitting, not exact symmetry
characters of the relativistic Bloch Hamiltonian. Likewise, changes
in positive band separations are not sufficient to identify the sign
or winding of a projected transverse mass.}

\section{Magnetic-orientation dependence of the relativistic spectrum}
\label{sec:rotation}

\rev{The final comparison isolates the role of the orientation of the
coplanar SVC relative to the crystal. We consider the reference texture
and coherent in-plane moment rotations by $\phi=\pm22.5^\circ$.
For each orientation, the spectra are compared on the same paths and
referenced to the corresponding parent Fermi level. Let $A(q)$ and
$A_{xy}(q)$ be two paths interchanged by $k_x\leftrightarrow k_y$.
For the tracked band pair define}
\begin{align}
\delta_\Delta(\phi,q)&=\Delta_A(\phi,q)-\Delta_{A_{xy}}(\phi,q),\\
\delta_{\bar E}(\phi,q)&=\bar E_A(\phi,q)-\bar E_{A_{xy}}(\phi,q),
\label{eq:Smirrorchannels}
\end{align}
\rev{where $\bar E$ denotes the band-pair midpoint. The midpoint
comparison probes an independent spectral channel rather than a quantity
inferred from the splitting.}

\begin{figure}[t]
\centering
\includegraphics[width=0.98\textwidth]{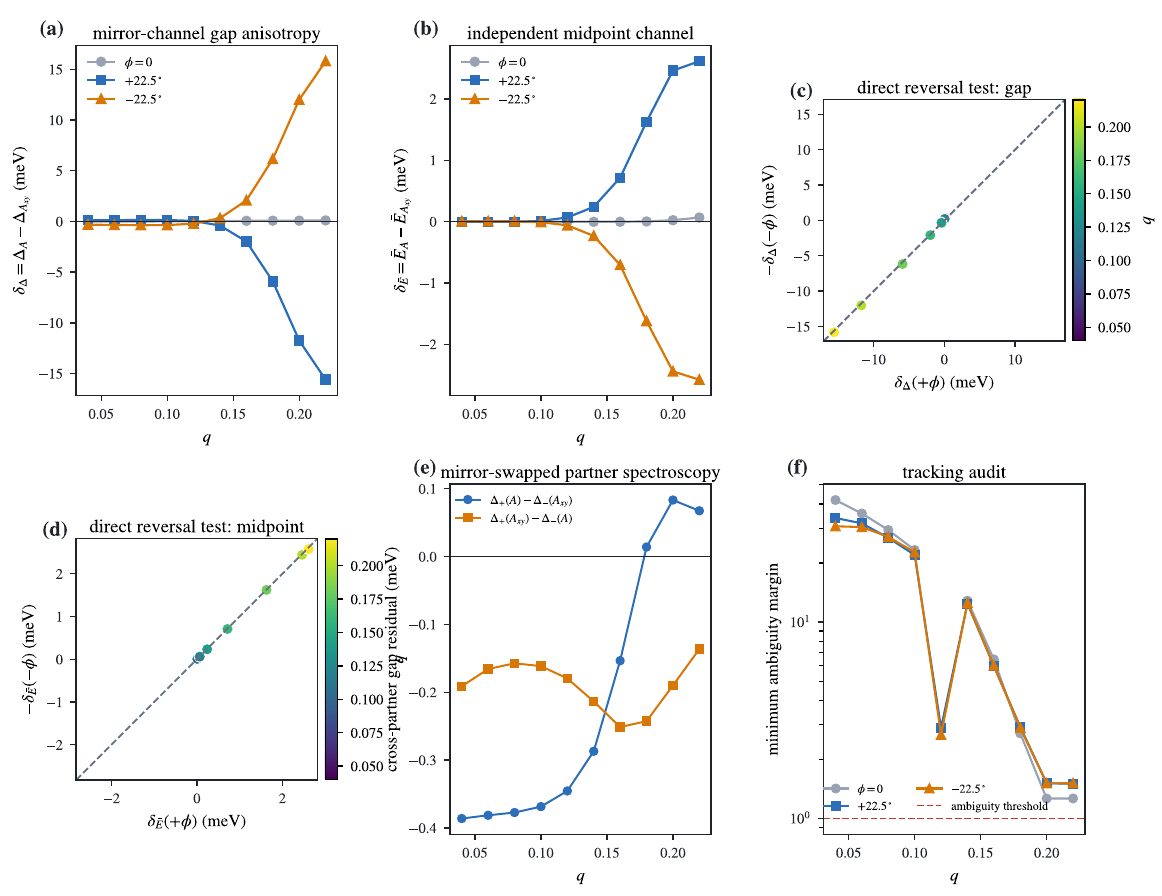}
\caption{\rev{\textbf{Opposite spin-rotation controls.}
(a) Mirror-channel splitting difference for the reference SVC and
rotations $\phi=\pm22.5^\circ$. (b) Corresponding difference of the
tracked band-pair midpoint. (c,d) Direct tests of the opposite-rotation
relation for both independent spectral channels.
(e) Comparison of mirror-interchanged partner paths.
(f) Band-tracking margins for the three textures.
The calculations establish reversal of spectral anisotropy under
opposite imposed rotations.}}
\label{fig:S7}
\end{figure}

\rev{The unrotated texture has an RMS channel difference of only
$0.057$~meV, whereas the two rotated configurations give $6.48$ and
$6.61$~meV. To quantify the response odd under reversal of $\phi$ we use}
\begin{align}
A_{\mathrm{odd}}^\Delta&=\frac12\mathrm{RMS}
 [\delta_\Delta(+\phi)-\delta_\Delta(-\phi)],\\
R_{\mathrm{odd}}^\Delta&=\mathrm{RMS}
 [\delta_\Delta(+\phi)+\delta_\Delta(-\phi)].
\end{align}
\rev{For the present data $A_{\mathrm{odd}}^\Delta=6.55$~meV and
$R_{\mathrm{odd}}^\Delta=0.20$~meV, corresponding to a residual of about
$3.1\%$ of the odd-response amplitude. The opposite-rotation profiles
are strongly anticorrelated: comparing $\delta_\Delta(+\phi)$ with
$-\delta_\Delta(-\phi)$ gives a correlation above $0.999$.
The band-pair midpoint behaves similarly, with an odd-response residual
of about $1.1\%$ of its amplitude. Directly interchanging the two paths
between opposite rotations and repeating the band-character tracking
provide independent consistency checks [Fig.~\ref{fig:S7}].}

\rev{These results show that the relativistic spectral anisotropy can be
reversed by an imposed reversal of the coherent moment rotation. The
positive gaps and band-pair midpoints characterize the electronic
reconstruction. They do not, by themselves, define the phase of a
complex mass pair or the associated Chern invariant. }

\bibliography{references}